\documentclass[pdflatex,sn-mathphys-num]{sn-jnl}

\usepackage[utf8]{inputenc}
\usepackage{textcomp}
\usepackage{newunicodechar}
\newunicodechar{ }{\,}
\usepackage{graphicx}%
\usepackage{epstopdf}
\usepackage{subcaption}
\usepackage{multirow}%
\usepackage{amsmath,amssymb,amsfonts}%
\usepackage{amsthm}%
\usepackage{mathrsfs}%
\usepackage[title]{appendix}%
\usepackage{xcolor}%
\usepackage{textcomp}%
\usepackage{manyfoot}%
\usepackage{booktabs}%
\usepackage{algorithm}%
\usepackage{algorithmicx}%
\usepackage{algpseudocode}%
\usepackage{listings}%
\definecolor{neRed}{RGB}{220,0,0}
\definecolor{neGreen}{RGB}{0,150,0}
\definecolor{neBlue}{RGB}{0,90,200}
\definecolor{neCyan}{RGB}{0,180,180}
\definecolor{neMagenta}{RGB}{200,0,150}
\definecolor{neYellow}{RGB}{240,200,0}
\definecolor{neDarkYellow}{RGB}{180,140,0}
\definecolor{neNavy}{RGB}{0,0,120}
\definecolor{nePurple}{RGB}{120,0,180}
\definecolor{neWine}{RGB}{150,0,60}
\definecolor{neOlive}{RGB}{90,110,40}
\definecolor{neOrange}{RGB}{230,120,20}

\theoremstyle{thmstyleone}%
\theoremstyle{thmstyletwo}%

\theoremstyle{thmstylethree}%

\begin{document}

\title[Low-Temperature Heat Capacity of Single-Walled Carbon Nanotube Bundles with low dimensional gas systems]{Low-Temperature Heat Capacity of Single-Walled Carbon Nanotube Bundles with low dimensional gas systems}


\author*[1]{\fnm{M.S.} \sur{Barabashko}}\email{barabaschko@ilt.kharkov.ua; ORCID: https://orcid.org/0000-0003-3168-7119}

\author[1]{\fnm{A.I.} \sur{Krivchikov}}\email{krivchikov@ilt.kharkov.ua; ORCID: https://orcid.org/0000-0001-5375-439X}

\affil[1]{\orgdiv{}\orgname{B. Verkin Institute for Low Temperature Physics and Engineering of the National Academy of Sciences of Ukraine}, \orgaddress{\street{47 Nauky Ave.}, \city{Kharkiv}, \postcode{61103}, \country{Ukraine}}}


\abstract{Low-temperature heat-capacity data for pristine single-walled carbon nanotube (SWCNT) bundles and for SWCNT/gas systems containing adsorbed Xe, Ne, N$_2$, and CH$_4$ were analyzed within a phenomenological framework.
The data were fitted within the phenomenological form $C(T)=C_{3}T^{3}+C_{5}T^{5}$, where the coefficients are interpreted as effective parameters of the coupled SWCNT--adsorbate excitation spectrum. Pristine SWCNT bundles exhibit a negative $C_{5}$ coefficient, reflecting a non-Debye dispersive contribution associated with low-energy flexural-like modes. Gas adsorption systematically modifies both $C_{3}$ and $C_{5}$, indicating that the adsorbate's heat-capacity contribution may not be treated as fully independent. The Xe series suggests that mass loading, modified adsorbate–nanotube interaction, and changes in intertube coupling slightly reshape the nanotube vibrational response. Coverage-dependent measurements for Ne reveal a progressive shift from nanotube-dominated excitations toward domination of additional low-energy adsorbate modes in the phonon spectrum of the SWCNT-Ne system, while molecular adsorbates (N$_2$, CH$_4$) introduce translational, librational, orientational, and rotational degrees of freedom that generate correspondingly stronger perturbations.}





\maketitle

\section{Introduction}\label{sec1}

Carbon nanotubes (CNTs) continue to attract significant attention due to their potential in hydrogen storage, hydrogen isotope separation, nanoelectronics, and gas-sensing technologies~\cite{kowalczyk2008cryogenic,barabashko2023influence,sun2024separation,barabashko2023proportional, krivchikov2026impact}. Among these applications, gas sensing and hydrogen storage are particularly sensitive to adsorption-induced modifications of the structural and vibrational properties of CNTs. This sensitivity has motivated extensive experimental and theoretical investigations of gas-adsorption phenomena in CNT systems~\cite{manzhelii2025localized,vinnikov2023hydrogen,danilchenko2014tunneling,niechcial2021separation,barabashko2023proportional,barabashko2023experimental,dolbin2009radial,dolbin2010low,faginas2021confinement}. The extent to which adsorbates modify CNT phonons depends strongly on the mass of the adsorbed particles, their coverage, and the adsorption site—factors that primarily govern vibrational behavior and low-temperature thermodynamics.

Single-walled carbon nanotubes (SWCNTs) are quasi-one-dimensional cylindrical nanostructures formed by seamlessly rolling a single graphene sheet into a hollow tube with nanometer-scale diameters and micrometer-scale lengths, resulting in exceptionally high aspect ratios~\cite{dresselhaus2004unusual}. The atomic structure and vibrational density of states of SWCNTs depend on the pair of chiral indices $(n,m)$, which specify the chiral vector on the graphene lattice and determine the nanotube geometry~\cite{cao2003exact}. Three structural classes arise from these indices—armchair $(n=m)$, zigzag $(m=0)$, and chiral $(n\neq m\neq 0)$—and the chirality directly governs whether a nanotube exhibits metallic or semiconducting electronic behavior~\cite{dresselhaus1998physical}. The diameter and chirality distributions of SWCNTs depend strongly on the synthesis growth method and the processing conditions, which also influence the extent to which individual nanotubes aggregate into bundles through van der Waals interactions, forming close-packed arrays containing tens to hundreds of tubes~\cite{ding2009dislocation, wu2022temperature}. Real materials inevitably contain structural imperfections—topological defects~\cite{barabashko2024kinetics}, local curvature variations~\cite{yildiz2021graphene}, and surface distortions—that further modulate the phonon density of states, adsorption energetics, and transport properties~\cite{barabashko2026universal,bagatskii2025low,balandin2011thermal}, and thus play a major role in low-temperature phonon physics.

In bundled SWCNTs, gas adsorption occurs at several distinct sites whose geometry and binding energies differ markedly~\cite{ulbricht2002desorption,zambano2001binding,gatica2001quasi}. The primary adsorption locations in closed-cap nanotubes include the interstitial channels (IC) between adjacent tubes, the grooves (G) at the bundle perimeter, and the outer surface (OS) of the bundle. These sites accommodate physisorbed atoms and molecules with site-specific confinement and interaction strengths. Adsorption inside closed-cap SWCNTs is inaccessible to large rare-gas atoms due to the absence of open tube ends and the constraints imposed by the cap structure~\cite{vinnikov2023hydrogen}. At low coverage, OS adsorption is negligible until the groove sites are filled. 
The caps of SWCNTs themselves consist of fullerene-like segments containing pentagons and hexagons that smoothly terminate the cylindrical lattice~\cite{dresselhaus1992carbon}. The sketch illustrates an idealized bundle with uniform tube diameters and defect-free outer surfaces. In reality, SWCNT bundles are not perfectly hexagonally packed, and bulk samples often contain voids, misaligned nanotubes, and non-uniform intertube spacing. Real bundles also exhibit diameter distributions, defect densities that depend on preparation conditions~\cite{barabashko2021experimental, barabashko2025experimental}, and partial opening of caps due to oxidation~\cite{dolbin2010low}. These factors strongly affect adsorption-site geometry and energetics.


Low-temperature thermodynamic measurements have revealed that the phonon spectrum of SWCNT bundles exhibits unique features associated with bundle coupling, aspect ratio, tube--tube van der Waals interactions, and the suppression of free-tube flexural modes. Bagatskii \textit{et al.}~\cite{bagatskii2012specific} investigated the specific heat at constant pressure, $C(T)$, of closed-end SWCNT bundles and, using combined experimental heat-capacity and thermal-expansion data, determined the Grüneisen parameter $\gamma(T)$. Their analysis showed that $\gamma(T)$ becomes negative below approximately 6~K, indicating the dominance of low-energy vibrational modes with negative Grüneisen parameters in this temperature regime. Recently, Barabashko \textit{et al.}~\cite{barabashko2025experimental} demonstrated that the heat capacity of SWCNT bundles below 3~K fitted within the phenomenological equation with two terms: a Debye term, $C_{3}T^{3}$, and a dispersive term, $C_{5}T^{5}$. The negative value of the $C_{5}$ coefficient provides evidence for flexural dispersion associated with out-of-plane ZA phonons~\cite{barabashko2025experimental,kosevich2006crystal,krivchikov2025impact, krivchikov2026impact}. Although compacted SWCNT bundles exhibit a negative $C_{5}T^{5}$ term, this term is generally used only as an empirical low‑temperature marker of non‑Debye dispersive behavior. In pristine SWCNTs, its sign and magnitude are consistent with the expected behaviour of flexural‑like low‑energy excitations. However, it does not provide a unique, mode‑resolved measure of a ZA branch for individual nanotube due to the interactions between nanotubes within the bundle.

The experimental heat capacity of SWCNT bundles with adsorbed gas impurities (Xe, Ne, N$_2$, CH$_4$) has been analyzed within models that treat the adsorbate as one-dimensional chains located along the grooves~\cite{barabashko2021experimental,syvokon2023collective,manzhelii2019phonon,siber2002phonons,kostov2003phonons,ramachandran2007thermodynamic}. At low coverage, the adsorbed species predominantly occupy the energetically most favorable adsorption sites in the groove (G) regions of closed-cap SWCNT bundles~\cite{bagatskii2013experimental,bagatskii2013heat, ulbricht2002desorption, dolbin2009radial, dolbin2010low}. The assumption of an additive heat-capacity contribution, $C_V(T)$, from Xe adsorbates on SWCNT bundles shows good agreement with theoretical calculations for one-dimensional chains of Xe atoms~\cite{bagatskii2013experimental, bagatskii2013heat,siber2002phonons, manzhelii2019phonon,barabashko2021experimental, syvokon2023collective, chishko2016heat}. Previous studies established that the IC, G, and OS sites are characterized by distinct binding energies~\cite{ulbricht2002desorption,siber2002phonons, manzhelii2019phonon,gatica2001quasi, zambano2001binding, chishko2016heat, ramachandran2007thermodynamic}. This diversity of adsorption environments significantly complicates both the theoretical description and the interpretation of experimental heat-capacity data for SWCNTs with large quantities of adsorbed gas atoms~\cite{ramachandran2007thermodynamic}, particularly at intermediate (partial monolayer) and high coverages (complete monolayer and partial second-layer occupation on the bundle surface). In addition, at higher coverage, mass loading can modify the flexural modes of SWCNTs, intertube coupling, and the effective stiffness of the SWCNT bundle. Therefore, the additivity approximation ceases to be valid once the interactions between the constituent subsystems become sufficiently strong to modify their individual contributions to the total response. Even at low coverage, the specific nature of the adsorbed species can significantly affect the heat-capacity behavior of particles in the grooves of SWCNTs: N$_2$ and CH$_4$ possess rotational degrees of freedom~\cite{bagatskii2013heat,bagatskii2016heat, barabashko2021experimental}, while H$_2$ and He exhibit pronounced zero-point motion and quantum delocalization~\cite{barabashko2026universal,pearce2005one}, which strongly influence adsorbate thermodynamics~\cite{wilson2003fiz,Lasjaunias2003}.

Despite substantial progress, a quantitative understanding of how adsorbates interact with structurally realistic SWCNT bundles---characterized by specific chirality distributions, defect populations, and adsorption-site energetics---remains incomplete. This gap is particularly relevant for cryogenic gas adsorption and isotope-selective processes, where quantum effects and confinement play decisive roles. Addressing these unresolved questions is essential for advancing CNT-based hydrogen technologies and for interpreting low-temperature thermodynamic measurements in nanotube systems.

Although the low-temperature heat capacity of pristine SWCNT bundles has been successfully described using a combination of a Debye term $C_{3}T^{3}$ and a dispersive flexural-phonon term $C_{5}T^{5}$, this framework has never been systematically applied to the effective coefficient $C_{3}$ and $C_{5}$, for SWCNT--gas systems. Prior studies of adsorbed gases on nanotube bundles have primarily modeled the adsorbate as one-dimensional chains or quasi-1D lattices confined to grooves or interstitial channels, without considering how adsorption modifies the intrinsic phonon spectrum of the nanotube matrix itself~\cite{barabashko2021experimental,syvokon2023collective,manzhelii2019phonon,siber2002phonons,kostov2003phonons,ramachandran2007thermodynamic, Lasjaunias2003,pearce2005one,bagatskii2013experimental,bagatskii2013heat,bagatskii2016heat, bagatskii2017heat}. In particular, no previous work has examined whether the presence of adsorbates alters the balance between the $C_{3}$ and $C_{5}$ contributions, nor whether the flexural-mode coefficient $C_{5}$---which encodes the dispersion of out-of-plane ZA phonons---responds systematically to the type, mass, or concentration of the adsorbed species. This gap is significant because adsorption perturbs the nanotube lattice in several coupled ways. Adsorbed atoms increase the effective mass per unit length of the tubes, modify local curvature at groove and surface sites, and change the van der Waals coupling. These perturbations alter the boundary conditions and stiffness of the nanotube network, which may in turn reshape the low‑energy phonon spectrum—especially the flexural (ZA) modes that dominate the heat capacity below a few kelvin.

The aim of this work is to investigate the $C_{3}$ and $C_{5}$ phonon-dispersion terms in SWCNT--gas systems and to establish how these contributions are modified by atomic and molecular adsorption. For the first time, we provide a quantitative assessment of how adsorbed species reshape the flexural-phonon landscape of SWCNT bundles and, consequently, the low-temperature thermodynamics of these systems. Specifically, we determine how adsorbed impurities influence the temperature dependence of the heat capacity of SWCNT bundles and evaluate how the flexural-mode coefficient $C_{5}$ varies with the type and concentration of the adsorbed atoms or molecules.

The additive reconstruction is used solely to place literature datasets
on a common thermodynamic basis. It does not by itself establish dynamical inde-
pendence of the nanotube and adsorbate subsystems, nor can deviations of the fitted
effective coefficients from those of pristine SWCNT bundles be uniquely assigned to
a renormalization of intrinsic nanotube phonons. Rather, the fitted coefficients char-
acterize the total low-energy response represented by the reconstructed SWCNT–
adsorbate system. Their systematic dependence on adsorbate identity and coverage motivates the search for a contribution from adsorption-induced modification of the
SWCNT vibrational spectrum.

\section{Brief characteristics of analyzed SWCNT bundles with adsorbed gases}\label{sec2}

For the heat-capacity analysis, we used experimental datasets for bundles of single-walled carbon nanotubes (SWCNTs) with adsorbed atomic and molecular gases. The systems include atomic adsorbates---xenon (Xe) \cite{bagatskii2013experimental} and neon (Ne) \cite{ramachandran2007thermodynamic}---and molecular adsorbates---nitrogen (N$_2$) \cite{bagatskii2013heat} and methane (CH$_4$) \cite{bagatskii2016heat,bagatskii2017heat}. The SWCNT bundle dataset \cite{bagatskii2012specific} serves as the reference, representing the intrinsic heat capacity of SWCNT bundles without adsorbed gas, whereas the gas-loaded datasets allow us to quantify adsorption-induced modifications of the heat capacity.

The experimental datasets provide a consistent basis for examining the low-temperature heat capacity of SWCNT bundles with different adsorbed gases. The Xe-, N$_2$-, and CH$_4$-loaded specimens were studied using closely related compacted c-SWCNT bundles with comparable nanotube diameters and controlled adsorbate loadings, whereas the Ne measurements offer a complementary dataset obtained for HiPCo$^{\mathrm{TM}}$ SWCNT bundles over a much broader range of adsorbate concentrations. Together, these four systems supply mutually reinforcing information on how adsorbate species, loading level, and adsorption conditions influence the low-energy thermodynamics of SWCNT bundles.

The adsorbates differ substantially in their physical characteristics: Xe and Ne are atomic species with markedly different masses, while N$_2$ and CH$_4$ are molecular adsorbates possessing additional internal degrees of freedom. Moreover, the Ne dataset spans  distinct loading levels, providing a unique opportunity to investigate the concentration dependence of the heat capacity. Taken collectively, these datasets enable us to assess whether adsorption-induced changes in the low-temperature heat capacity can be described primarily by an additive contribution from the adsorbate subsystem, or whether adsorption also modifies the low-energy vibrational spectrum of the SWCNT host.

Particular attention is devoted to the low-temperature representation:

\begin{equation}
C(T) = C_3 T^3 + C_5 T^5
\label{eq:heat_fit}
\end{equation}

where $C_3$ captures the leading acoustic (Debye-like) contribution and $C_5$ reflects the effective dispersive term associated with the low-energy flexural phonon spectrum. Comparing pristine and gas-loaded samples enables us to determine whether adsorption manifests primarily as an additional heat-capacity contribution from the adsorbate subsystem or whether it is accompanied by a measurable modification of the flexural-phonon component intrinsic to the SWCNT bundle. The evolution of the effective $C_{3}$ and $C_{5}$ coefficients with adsorbate type and loading provides a quantitative basis for assessing possible changes in the flexural-phonon contribution and for testing the applicability of the proposed phenomenological approximation.

It is important to distinguish this phenomenological interpretation from a direct 
mode-by-mode decomposition of the experimental heat capacity. The coefficients 
$C_{3}$ and $C_{5}$ describe the net low-temperature response of the bundle 
and may include the effects of intertube coupling, structural heterogeneity, and the 
distribution of nanotube geometries present in the experimental sample. In particular, 
the reference sample is a compacted bundle material rather than an isolated SWCNT. 
The resulting coefficients should therefore be regarded as effective parameters 
characterizing the collective low-energy excitations of SWCNT bundles.

Although the coefficients $C_{3}$ and $C_{5}$ are obtained from a 
phenomenological fit, it is important to clarify the physical origin of the $T^{5}$ term. 
A pure quadratic flexural branch, as found in isolated SWCNTs or strictly 1D/2D 
systems, produces low-temperature heat capacities scaling only as $T^{1/2}$ (1D) \cite{gospodarev2017phonon,popov2004theoretical} or 
$T$ (2D) \cite{gospodarev2017phonon,krivchikov2025impact,krivchikov2026impact}. Such dispersions therefore would not be expected to generate a $T^{5}$ contribution. The 
appearance of the $T^{5}$ term in compacted SWCNT bundles arises only because real 
samples consist of mechanically coupled nanotubes. Intertube interactions, residual 
tension, and adsorbate-induced pinning modify the long-wavelength flexural spectrum 
so that its behavior is no longer purely quadratic. Instead, the ZA branch acquires a 
small linear component at very low energies, similar to the situation described by 
Lifshitz and Kosevich for weakly coupled or stretched chains. In such 3D-coupled 
systems, the low-energy phonons undergo a crossover from an acoustic-like regime at 
the lowest frequencies to a flexural regime at higher frequencies, and the next-to-leading 
term in the low-temperature expansion of the heat capacity is precisely the $T^{5}$ 
contribution.

Thus, the fitted coefficients $C_{3}$ and $C_{5}$ should be interpreted as 
effective parameters describing the collective excitations of the bundle, and the 
presence of the $T^{5}$ term reflects the linear–to–flexural crossover that appears only 
when SWCNTs are coupled into a three-dimensional bundle.

This configuration provides a robust framework for analyzing the contribution of adsorbed atoms and molecules to the total heat capacity of the SWCNTs/gas system, including the formation of quasi-one-dimensional adsorbate chains, changes in intertube coupling, and modulation of flexural phonon modes with increasing coverage.

\subsection{SWCNT bundles with adsorbed Xe}\label{subsec2}

For the analysis, we use the experimental heat-capacity dataset of SWCNT bundles with adsorbed Xe~\cite{bagatskii2013experimental}, together with the corresponding dataset for pristine SWCNT bundles~\cite{bagatskii2012specific}. These datasets provide the necessary basis for decomposing the total heat capacity according to Eq.~\ref{eq:heat_fit} into the Debye contribution $C_3 T^3$ and the flexural-phonon contribution $C_5 T^5$, and for analyzing the impact of adsorbed gas additives on the heat capacity of the SWCNTs--Xe system.

Bagatskii et al.~\cite{bagatskii2013experimental} performed low-temperature calorimetric measurements on bulk samples of closed-end single-walled carbon nanotube (c-SWCNT) bundles containing adsorbed xenon. The SWCNT powder, synthesized by catalytic chemical vapor deposition and consisting of more than 90\% SWCNT bundles with an average tube diameter of approximately 1.1\,nm, was compacted at 1.1\,GPa to form a dense bulk specimen with partial nanotube alignment perpendicular to the compression axis~\cite{bagatskii2012specific}. The compacted sample mass was $716.00 \pm 0.05$\,mg, and the same specimen was used for both pristine and Xe-loaded measurements, ensuring direct comparability of the heat-capacity data. Xenon loading amounted to $(3.19 \pm 0.05)\times10^{-4}$\,mol ($\approx 7.15$\,scc), corresponding to an adsorbate-to-carbon ratio of $N_{\mathrm{Xe}}/N_{\mathrm{C}} \approx 0.55\%$. Here $1\,\mathrm{scc}$ (standard cubic centimeter at STP) equals $4.46\times10^{-5}$\,mol. The Xe purity was 99.98\%, with trace $\mathrm{N}_2$ and Kr impurities, and the adsorbate content was determined using a standard PVT procedure. The heat capacity $C(T)$ of the Xe-loaded sample was obtained after subtraction of the calorimeter addenda. At the temperatures considered, the difference between $C_P$ and $C_V$ for SWCNT bundles is negligible, allowing the measured quantity to be treated simply as $C(T)$~\cite{bagatskii2012specific}. This well-defined low-coverage Xe/SWCNT system provides a reliable basis for analyzing adsorption-induced modifications of the nanotube heat capacity.

\subsection{SWCNT bundles with adsorbed Ne}\label{subsec2}

Ramachandran and Vilches~\cite{ramachandran2007thermodynamic} carried out low-temperature calorimetric measurements on bundles of single-walled carbon nanotubes (SWCNTs) containing adsorbed neon in the temperature range 1.8--19\,K. The sample consisted of 0.413\,g of HiPCo$^{\mathrm{TM}}$ SWCNT bundles (Carbon Nanotechnologies Incorporated), synthesized via the high-pressure CO disproportionation (HiPCo$^{\mathrm{TM}}$) process and containing approximately 3\,wt.\% Fe as residual catalyst. The mean nanotube diameter of the HiPCo$^{\mathrm{TM}}$ material was $\approx 1.1$\,nm~\cite{chiang2001purification}, comparable to that of the SWCNT bundles examined by Bagatskii~\cite{bagatskii2012specific,bagatskii2013heat,bagatskii2013experimental,bagatskii2016heat,bagatskii2017heat}. Neon loadings were investigated, from 1.99 to 153.6\,scc adsorbed by the SWCNT bundles (see Table~\ref{tab:fit_Cp}). Reconstructed heat‑capacity data for the SWCNTs–Ne system were obtained using a mixing rule applied to the reference heat capacity of pristine SWCNTs measured by Bagatskii \cite{bagatskii2012specific} and the heat capacity of Ne adsorbates measured by Ramachandran and Vilches~\cite{ramachandran2007thermodynamic}. This dataset provides a detailed experimental foundation for assessing how Ne adsorption modifies the low-temperature heat capacity of SWCNT bundles and for comparing the Ne-loaded system with the Xe, N$_2$, and CH$_4$ specimens analyzed in the present study.

\subsection{SWCNT bundles with adsorbed N$_2$}\label{subsec3}

Bagatskii et al.~\cite{bagatskii2013heat} extended their low-temperature calorimetric investigations to closed-end single-walled carbon nanotube (c-SWCNT) bundles saturated with nitrogen, using the same adiabatic calorimeter~\cite{bagatskii2011simple} and the same compacted bulk SWCNT sample previously characterized in measurements of pristine bundles~\cite{bagatskii2012specific}. This experimental continuity ensures that the effects of N$_2$ adsorption can be evaluated directly relative to the heat capacity of an identical SWCNT host. The amount of adsorbed nitrogen was $(3.81 \pm 0.06)\times10^{-4}$\,mol, corresponding to approximately $8.54$\,scc, where $1\,\mathrm{scc}$ (standard cubic centimeter at STP) equals $4.46\times10^{-5}$\,mol. The nitrogen purity was 99.997\%, with O$_2$ at 0.003\% as the only detected impurity. Saturation of the c-SWCNT bundles was carried out directly in the vacuum chamber of the calorimeter by cooling the calorimeter cell~\cite{bagatskii2011simple}, ensuring controlled adsorption under equilibrium conditions. From the measured nitrogen uptake and the known carbon content of the compacted sample, Bagatskii et al.\ determined an ratio of $N_{\mathrm{N_2}}/N_{\mathrm{C}} = 0.0066$, which defines the loading level analyzed here. Together with the corresponding Xe- and CH$_4$-loaded samples, this N$_2$-loaded system provides a basis for assessing whether the low-temperature heat capacity of SWCNT bundles exhibits systematic dependence on the mass, molecular character, or internal degrees of freedom of the adsorbed species; in particular, N$_2$ is noteworthy because, unlike atomic Xe, it possesses rotational modes that may contribute to the thermodynamics of the adsorbed phase.

\subsection{SWCNT bundles with adsorbed CH$_4$}\label{subsec4}

In the present work, we analyze the heat capacity of the SWCNT--CH$_4$ system using the same methodological framework applied to the Xe-, Ne-, and N$_2$-loaded specimens, thereby ensuring a consistent basis for evaluating how adsorbed methane influences the low-temperature heat capacity of SWCNT bundles.

Bagatskii et al.~\cite{bagatskii2016heat} investigated the heat capacity of closed-end single-walled carbon nanotube (c-SWCNT) bundles saturated with methane. The amount of CH$_4$ introduced into the calorimeter vacuum chamber was $(2.018 \pm 0.06)\times10^{-4}$\,mol, selected using the same geometric model previously applied to determine the quantities of Xe and N$_2$ required to fill the external groove sites of the bundles. This amount corresponds to approximately $4.53$\,scc of methane, where $1\,\mathrm{scc}$ (standard cubic centimeter at STP) equals $4.46\times10^{-5}$\,mol. The methane quantity was determined using the standard PVT method. From the known carbon content of the compacted SWCNT sample, Bagatskii et al.\ calculated an adsorbate-to-carbon atomic ratio of $N_{\mathrm{CH_4}}/N_{\mathrm{C}} \approx 0.0034$. The chemical purity of CH$_4$ was 99.96\,\%, with 0.03\,\% N$_2$ as the only detectable impurity~\cite{bagatskii2017heat}.

\section{Discussion of the Heat Capacity of SWCNTs–Gas Systems}\label{sec3}

The experimental datasets considered in this work were obtained using different calorimetric configurations and adsorption procedures. Consequently, the quantities reported in the original publications are not always directly comparable. In particular, the calorimetric measurements of Ramachandran and Vilches \cite{ramachandran2007thermodynamic} were performed on HiPCo™ SWCNT bundles mounted between thin-wall copper foils, and the reported quantity corresponds to the heat capacity of the calorimetric assembly containing the SWCNT–Ne system. In contrast, the measurements of Bagatskii et al. \cite{bagatskii2012specific,bagatskii2013experimental,bagatskii2013heat,bagatskii2016heat,bagatskii2017heat} were performed using an adiabatic calorimeter in which the contributions of the calorimeter addenda were determined separately \cite{bagatskii2011simple}. Moreover, the original studies did not all report the heat capacity of the SWCNT host and the adsorbed phase in the same normalization. A uniform reconstruction procedure is therefore required before the different datasets can be compared quantitatively.

\subsection{Uniform procedure for the Analysis of Low-Temperature Heat Capacity of SWCNT–Gas Systems }\label{subsec1}

A consistent reconstruction procedure is required because the available calorimetric datasets differ in how the heat capacity of the SWCNT--gas system was originally reported. In several studies, including that of Ramachandran and Vilches~\cite{ramachandran2007thermodynamic}, only the total heat capacity of the calorimeter cell containing SWCNT bundles mounted between thin copper foils was reported. Similar limitations occur in other calorimetry experiments~\cite{wilson2003fiz,Lasjaunias2003}, where only the heat capacity of the adsorbed film was reported. In the original studies, the adsorbate contribution was often treated within an additive approximation~\cite{bagatskii2013experimental,bagatskii2013heat,bagatskii2016heat,bagatskii2017heat,ramachandran2007thermodynamic,wilson2003fiz,Lasjaunias2003}. To place all systems on a common basis, we use the additive representation:
\begin{equation}
C_{\mathrm{tot}}(T) = C_{\mathrm{SWCNT}}(T) + C_{\mathrm{ads}}(T),
\label{eq:additive}
\end{equation}
where $C_{\mathrm{SWCNT}}(T)$ denotes the heat capacity of pristine SWCNT bundles~\cite{bagatskii2012specific} and $C_{\mathrm{ads}}(T)$ is the experimentally reported contribution of the adsorbed gas film~\cite{ramachandran2007thermodynamic,bagatskii2013experimental,bagatskii2013heat,bagatskii2016heat}. Equation~\ref{eq:additive} is used here as a reconstruction tool rather than as an assumption of complete dynamical independence between the nanotube and adsorbate subsystems; indeed, one objective of the present work is to test whether adsorption modifies the low-energy vibrational spectrum of the SWCNT host.

For the Xe-, N$_2$-, and CH$_4$-loaded systems, the corresponding pristine SWCNT heat capacity is available from measurements on compacted closed-end SWCNT bundles~\cite{bagatskii2012specific}, enabling a direct reconstruction of $C_{\mathrm{tot}}(T)$ from the reported adsorbate contribution. The Ne dataset requires a different treatment: Ramachandran and Vilches~\cite{ramachandran2007thermodynamic} reported only the total heat capacity of the SWCNT--Ne assembly. In this case, the reported experimental Ne contribution \cite{ramachandran2007thermodynamic} is combined with the independently measured pristine-SWCNTs heat capacity \cite{bagatskii2012specific} to place the SWCNTs--Ne system on the same reference basis as the Xe-, N$_2$-, and CH$_4$-containing samples. Additionally, the HiPCo$^{\mathrm{TM}}$ material in \cite{ramachandran2007thermodynamic} contains $\approx 3$\,wt.\% residual Fe catalyst and differs in synthesis route from the compacted c-SWCNTs bundles. Therefore, the Ne films dataset is treated as complementary rather than strictly equivalent to the Bagatskii datasets.

A distinction must be made between directly measured quantities and reconstructed quantities. Depending on the experiment, the reported heat capacity may correspond to the total heat capacity of the SWCNTs--gas system, the heat capacity of the adsorbed film after subtraction of a reference contribution, or a mass-normalized heat capacity. These quantities are not interchangeable without explicit normalization. All adsorbate loadings are therefore converted to standard cubic centimeters at STP,

\[
1\,\mathrm{scc} = 4.46\times10^{-5}\,\mathrm{mol},
\]

and subsequently to adsorbate mass $m_{\mathrm{gas}}$ using the molar mass of the species. Where available, the adsorbate-to-carbon ratio $N_{\mathrm{ads}}/N_{\mathrm{C}}$ is retained as a complementary measure of coverage.

When the heat capacity is expressed per unit mass of the complete SWCNT--gas sample, Eq.~\ref{eq:additive} becomes
\begin{equation}
c_{\mathrm{SWCNT\text{-}gas}}(T) =
\omega_{\mathrm{SWCNT}}\,c_{\mathrm{SWCNT}}(T) +
\omega_{\mathrm{gas}}\,c_{\mathrm{gas}}(T),
\label{eq:mixture}
\end{equation}
where $c_{\mathrm{SWCNT}}(T)$ and $c_{\mathrm{gas}}(T)$ are the specific heat capacities of the pristine SWCNT material and adsorbed gas, respectively, and the mass fractions satisfy $\omega_{\mathrm{SWCNT}} + \omega_{\mathrm{gas}} = 1$. Equation~\ref{eq:mixture} is applied only when the original data are mass-normalized; otherwise, the reported quantity first needs to be converted by the common normalization.

The reconstruction procedure therefore consists of: (i) identifying the normalization of the reported heat capacity; (ii) converting the adsorbate loading from scc to moles and, where required, to mass; (iii) introducing the pristine-SWCNT reference contribution; and (iv) expressing the reconstructed heat capacity on a common mass basis over the temperature range where the low-temperature phonon expansion is applicable. The resulting datasets for Xe, N$_2$, CH$_4$, and Ne provide a unified framework for comparing systems with different adsorbate masses, molecular structures, adsorption levels, and experimental histories. Importantly, the additive reconstruction defines only the baseline description; systematic deviations in the effective coefficients of the low-temperature $C_{3}T^{3} + C_{5}T^{5}$ representation serve as potential signatures of adsorption-induced modifications of the flexural phonon contribution and low-energy vibrational spectrum of SWCNT bundles.

\subsection{Quantitative analysis of the low-temperature heat capacity SWCNT-gas system}\label{subsec2}

The procedure described above provides a common basis for comparing the low-temperature heat capacities of SWCNT bundles containing different adsorbed species. The central objective of the present analysis is to determine whether the adsorption-induced changes can be described by an additive contribution from the adsorbate or whether they are accompanied by modifications of the total low-energy vibrational spectrum of the SWCNT--gas system.

Because the experimental datasets were obtained with different adsorbates as well as different SWCNT materials~\cite{bagatskii2012specific,ramachandran2007thermodynamic,Lasjaunias2003}, a direct comparison of the absolute heat capacities is not sufficient to distinguish these two possibilities. We therefore analyze the data in terms of the low-temperature expansion established for pristine SWCNT bundles (Equation~\ref{eq:heat_fit}) and use the pristine-SWCNT heat capacity~\cite{bagatskii2012specific,barabashko2025experimental} as a reference. The coefficient $C_{3}$ characterizes the leading acoustic contribution, whereas $C_{5}$ describes the first dispersive correction to the low-energy vibrational spectrum. Of particular interest is the experimentally observed negative value of $C_{5}$, which has been associated with flexural dispersion of out-of-plane ZA modes~\cite{barabashko2025experimental}.

The analysis is performed in two steps. First, the heat capacity of pristine SWCNT bundles is established as a reference over the temperature interval in which Eq.~\ref{eq:heat_fit} provides an adequate description. This reference defines the intrinsic low-energy contribution of the nanotube host in the absence of adsorption. Second, the reconstructed heat capacities of the Xe-, N$_2$-, CH$_4$-, and Ne-containing systems are compared with this reference. The corresponding changes in $C_{3}$ and $C_{5}$ are then examined as functions of adsorbate species and loading.

The distinction between the two possible mechanisms is important. If adsorption is predominantly additive, the heat capacity of the SWCNT host should retain approximately the same $C_{3}$ and $C_{5}$ coefficients as the pristine material, while the additional heat capacity should be attributable to the adsorbate subsystem. In contrast, a systematic evolution of the effective $C_{3}$, and particularly $C_{5}$, coefficients with adsorption would indicate that the low-energy vibrational spectrum of the nanotube bundle is not invariant upon adsorption. Such changes could arise from modifications of the intertube interaction, effective bending rigidity, local mass distribution, or coupling between the nanotubes and the adsorbed species.

The Ne dataset is especially useful in this respect because it contains measurements for  different adsorption concentration. It therefore provides an opportunity to examine not only the difference between pristine and adsorbate-containing samples but also the evolution of the low-temperature heat capacity with increasing adsorbate loading. The Xe, N$_2$, and CH$_4$ systems provide an additional comparison between atomic and molecular adsorbates and allow the possible role of adsorbate mass and internal molecular degrees of freedom to be considered.

It should be emphasized that the coefficients obtained from fits to the adsorption data are treated as effective coefficients. They characterize the experimentally observed low-temperature dependence of the complete SWCNT--gas system and do not, by themselves, constitute a unique decomposition into host and adsorbate contributions. Their physical interpretation is therefore based on comparison with the pristine reference, on their systematic dependence on adsorbate loading, and on the differences between chemically and structurally distinct adsorbates. This approach allows us to test the additive description without imposing it as an \emph{a priori} physical constraint.

A reliable reference for the heat capacity of pristine SWCNT bundles is essential for identifying adsorption-induced changes in the low-energy thermodynamic response. In the present work, we use the low-temperature heat-capacity data for compacted closed-end SWCNT bundles reported by Bagatskii et al.~\cite{bagatskii2012specific}. The same material served as the host for the Xe-, N$_2$-, and CH$_4$-adsorption measurements, providing an important advantage: differences between the pristine and adsorbate-containing systems are unlikely to be attributable to variations in nanotube diameter, synthesis route, or sample preparation.

At temperatures below approximately 3\,K, the heat capacity of the pristine SWCNT bundles is well represented by
\begin{equation}
C^{(0)}(T) = C_{3}^{(0)}T^{3} + C_{5}^{(0)}T^{5},
\label{eq:pristine_fit}
\end{equation}
where the superscript $(0)$ denotes the pristine SWCNTs without gas (reference state). The leading $C_{3}T^{3}$ term reflects the contribution of low-energy acoustic excitations, while the $C_{5}T^{5}$ term associated in pristine
SWCNT bundles with the low-energy flexural ZA phonons. In SWCNT bundles, this correction is particularly significant because the low-energy spectrum contains flexural modes whose dispersion deviates from that of an ideal three-dimensional Debye solid.

For pristine SWCNTs~\cite{bagatskii2012specific,barabashko2025experimental}, the fitted value of $C_{5}^{(0)}$  is negative (Table~\ref{tab:fit_Cp}). 
This behavior is consistent with the quadratic flexural character of the out-of-plane 
ZA acoustic dispersion and has been interpreted as the signature of the flexural 
contribution associated with ZA phonons\cite{barabashko2025experimental, kosevich2006crystal}. The negative $C_{5}^{(0)}$ therefore 
serves as a sensitive phenomenological indicator of the dispersive, non-Debye 
character of the low-energy vibrational spectrum of the pristine SWCNT bundle.

It is important to distinguish this phenomenological interpretation from a direct mode-by-mode decomposition of the experimental heat capacity. The coefficients $C_{3}^{(0)}$ and $C_{5}^{(0)}$ describe the net low-temperature response of the bundle and may include the effects of intertube coupling, structural heterogeneity, and the distribution of nanotube geometries present in the experimental sample. In particular, the reference sample is a compacted bundle material rather than an isolated individual SWCNTs. The resulting coefficients should therefore be regarded as effective parameters characterizing the collective low-energy excitations of the SWCNTs in the out-of-plane direction.

Equation~\ref{eq:pristine_fit} defines the baseline against which the adsorption data are analyzed. If adsorption were strictly additive and did not modify the host vibrational spectrum, the intrinsic SWCNT contribution would remain described by the same coefficients $C_{3}^{(0)}$ and $C_{5}^{(0)}$, and the additional heat capacity would originate solely from the adsorbed species:
\begin{equation}
C_{\mathrm{tot}}(T) = C_{\mathrm{SWCNT}}^{(0)}(T) + C_{\mathrm{ads}}(T).
\label{eq:additive_reference}
\end{equation}
Under this assumption, changes in the total heat capacity with increasing adsorption could be reproduced without altering the reference coefficients $C_{3}^{(0)}$ and $C_{5}^{(0)}$.

The present analysis does not impose this condition. Instead, Eq.~\ref{eq:additive_reference} is used as an additive baseline against which the experimental data are tested. For each adsorbate, the reconstructed low-temperature heat capacity is fitted independently, and the resulting effective coefficients are compared with $C_{3}^{(0)}$ and $C_{5}^{(0)}$. In particular, systematic changes in $C_{5}$ with adsorbate loading are of primary interest because they would suggest that the low-energy dispersive contribution may not be fully captured by an unchanged pristine-SWCNT spectrum plus an independent adsorbate heat capacity.

This reference-state analysis is therefore central to the subsequent interpretation. The pristine SWCNT data establish both the leading acoustic contribution and the characteristic negative dispersive correction against which all adsorption-induced changes can be quantified. The Xe-, N$_2$-, CH$_4$-, and Ne datasets can then be treated within the same phenomenological framework, allowing the effects of adsorbate mass, molecular structure, and loading to be separated from the intrinsic low-temperature response of the nanotube host.

\section{Discussion of flexural phonons in SWCNT bundles with adsorbed gases}\label{sec4}

\subsection{Low temperature heat capacity of SWCNT bundles with adsorbed Xe}\label{subsec1}

\begin{figure}[htbp]
  \centering
  \includegraphics[width=0.95\textwidth]{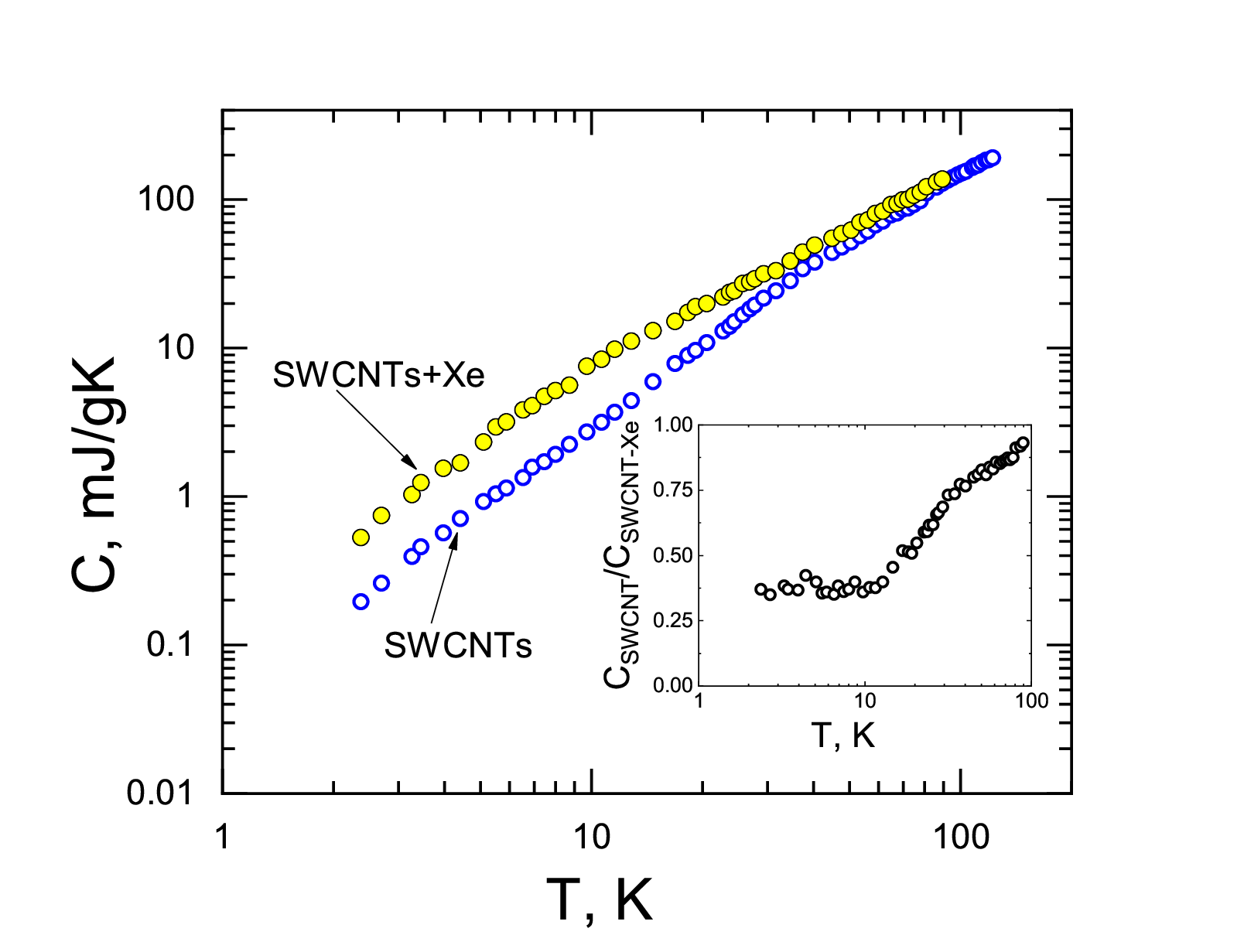}
 
  \caption{Temperature dependence of the heat capacity of pristine c-SWCNT bundles (\textcolor{blue}{\(\circ\)}) and c-SWCNT bundles with adsorbed xenon (\textcolor{yellow}{\(\bullet\)}). The heat capacity $C(T)$ of closed-end SWCNT bundles is shown together with that of the SWCNT--Xe system containing 7.15\,scc of adsorbed Xe. The inset displays the ratio $C_{\mathrm{SWCNT}}/C_{\mathrm{SWCNT\text{-}Xe}}$ \textcolor{black}{\(\circ\)}), highlighting the relative contribution of the nanotube matrix to the total heat capacity of the SWCNT--Xe system across the investigated temperature range.}
  \label{fig:fig_2}
\end{figure}

Figure~\ref{fig:fig_2} compares the total heat capacity of the SWCNT--Xe (7.15\,scc) \cite{bagatskii2013experimental} system with that of pristine compacted c-SWCNT bundles~\cite{bagatskii2012specific}. Xenon adsorption produces a pronounced enhancement of the heat capacity across the entire temperature range investigated, although the relative difference between the two systems decreases with increasing temperature. The inset of Fig.~\ref{fig:fig_2} shows the ratio $C_{\mathrm{SWCNT}}/C_{\mathrm{SWCNT\text{-}Xe}}$, which quantifies the fraction of the total heat capacity attributable to the nanotube host. Below approximately 10\,K, the Xe-containing sample exhibits a substantially larger heat capacity than the pristine bundles, indicating a significant contribution from low-energy excitations associated with the adsorbed Xe. In this regime, the motion of Xe atoms is strongly constrained by the adsorption potential, and the adsorbate behaves predominantly as a localized mass load coupled to the nanotube lattice. As the temperature increases, the relative contribution of the Xe subsystem diminishes, and the heat capacity of the nanotube framework becomes increasingly dominant. The change in local slope near 10 K for the $C_{\mathrm{SWCNT}}/C_{\mathrm{SWCNT\text{-}Xe}}$ reflects the thermal population of higher-energy phonon branches in the SWCNT bundles: long-wavelength out-of-plane flexural (ZA-type) modes govern the lowest-temperature behavior of low-dimensional carbon systems~\cite{barabashko2025experimental,krivchikov2025impact,krivchikov2026impact,kosevich2006crystal,gospodarev2017phonon}, whereas transverse acoustic (TA), longitudinal acoustic (LA), and low-lying optical modes become increasingly populated at elevated temperatures. Thermally activated redistribution of Xe atoms between groove and outer-surface adsorption sites becomes appreciable only above roughly 30\,K~\cite{barabashko2023experimental}, so the low-temperature data primarily probe the vibrational response of a SWCNT bundle with a relatively localized Xe mass loading.

\begin{figure}[htbp]
  \centering
  \begin{subfigure}{0.7\textwidth}
    \centering
    \includegraphics[width=\textwidth]{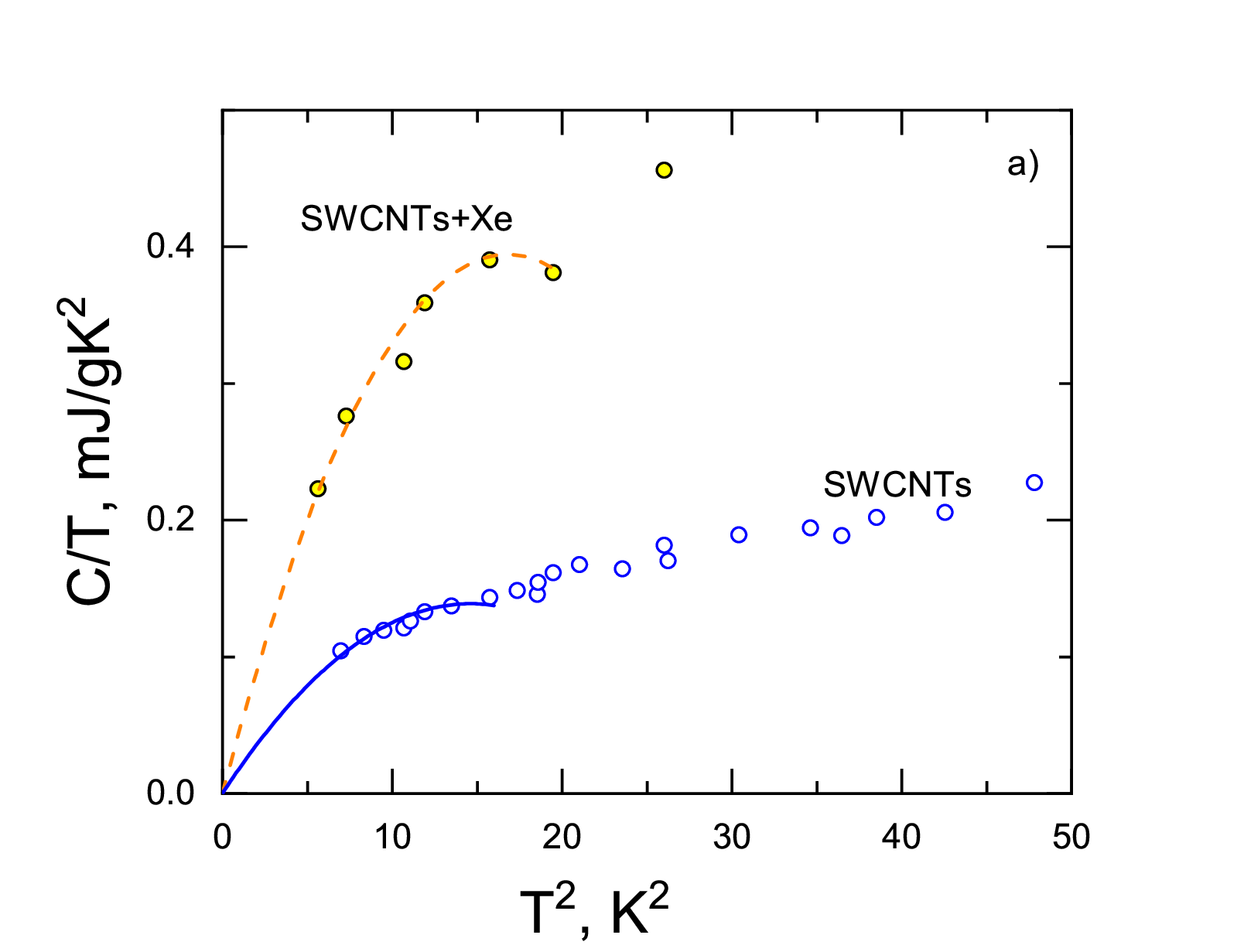}
    \caption{Temperature dependence of $C/T$ plotted as a function of $T^{2}$ for pristine c\mbox{-}SWCNT bundles (\textcolor{blue}{\(\circ\)}) and c\mbox{-}SWCNT bundles with adsorbed xenon (\textcolor{yellow}{\(\bullet\)}). The solid and dotted lines represent fits to the experimental data using Eq.~\ref{eq:heat_fit} for the pristine and Xe\mbox{-}containing SWCNT bundles, respectively.}
    \label{fig:fig_3a}
  \end{subfigure}

  \par\medskip

  \begin{subfigure}{0.7\textwidth}
    \centering
    \includegraphics[width=\textwidth]{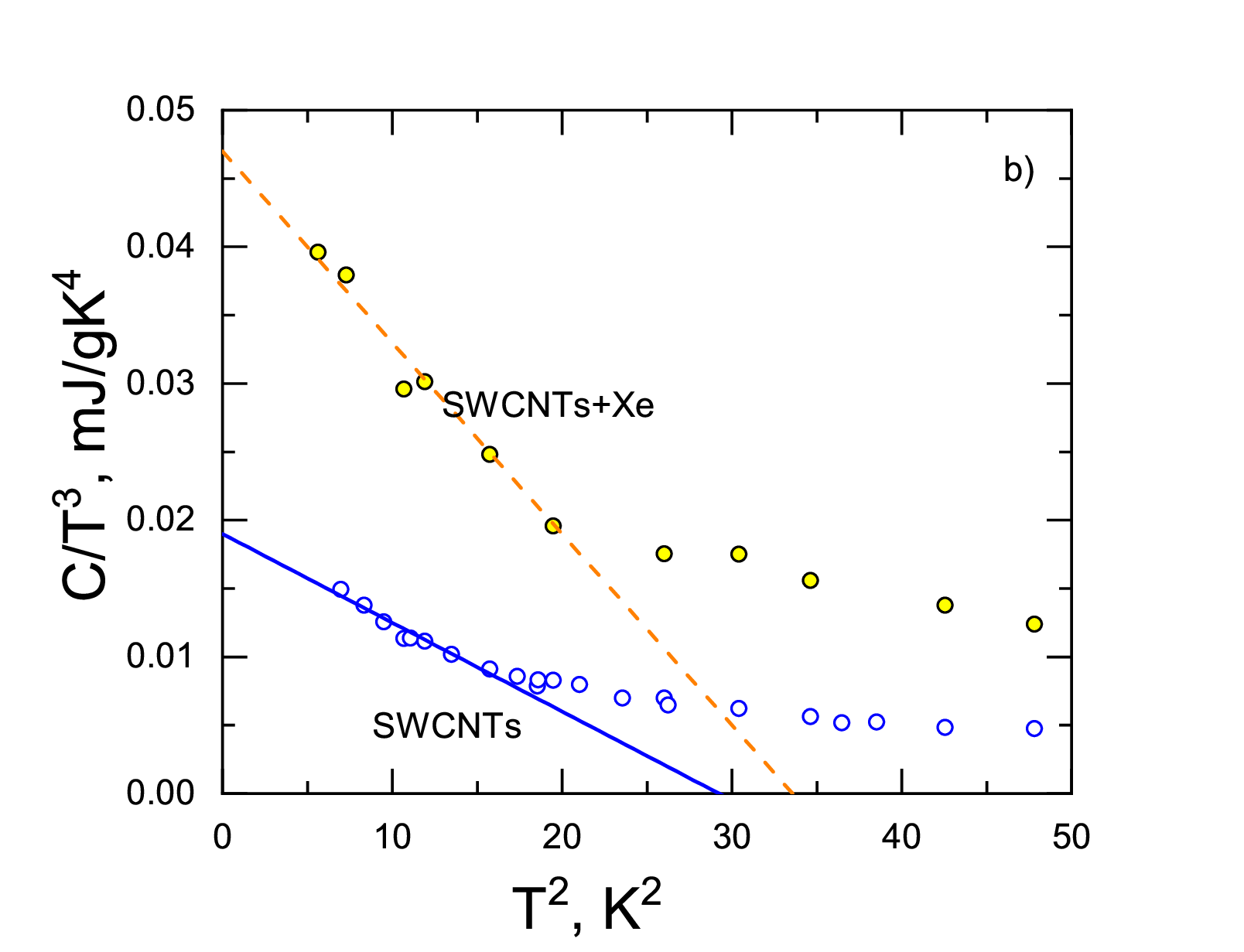}
    \caption{Temperature dependence of $C/T^{3}$ plotted as a function of $T^{2}$ for pristine c\mbox{-}SWCNT bundles (\textcolor{blue}{\(\circ\)}) and c\mbox{-}SWCNT bundles with adsorbed xenon (\textcolor{yellow}{\(\bullet\)}). The solid and dotted lines represent fits to the experimental data using Eq.~\ref{eq:heat_fit} for the pristine and Xe\mbox{-}containing SWCNT bundles, respectively.}
    \label{fig:fig_3b}
  \end{subfigure}

  \caption{Low-temperature representations of the heat capacity of pristine and Xe-loaded c\mbox{-}SWCNT bundles.}
  \label{fig:fig_3}
\end{figure}

The low-temperature heat capacities of pristine c-SWCNT bundles and the c-SWCNT--Xe system were examined below 7\,K using the standard $C/T$ versus $T^{2}$ representation (Fig.~\ref{fig:fig_3a}) and the complementary $C/T^{3}$ versus $T^{2}$ plot (Fig.~\ref{fig:fig_3b}). Both representations indicate that, for the SWCNT and SWCNT--Xe systems, the heat capacity below 4\,K is well described by the polynomial expansion Eq.~\ref{eq:heat_fit}: $C(T)=C_{3}T^{3}+C_{5}T^{5}$,  
which captures the dominant phonon contributions in this regime. Figures~\ref{fig:fig_3a} and Figure~\ref{fig:fig_3b} yield mutually consistent polynomial expansion Eq.~\ref{eq:heat_fit} for $C_{3}$ and $C_{5}$, supporting the robustness of the polynomial approximation in the sub-4\,K region. The observed changes in the fitted parameters upon Xe adsorption provide a sensitive probe of how weakly bound atomic species modify the low-energy vibrational spectrum of the SWCNT--Xe system. The extracted coefficients are summarized in Table~\ref{tab:fit_Cp}.  Xe adsorption increases the effective low-temperature heat capacity and changes both fitted coefficients.
Because Xe has no molecular rotational degrees of freedom, this result is consistent with an important
contribution from adsorbate translational dynamics and adsorption-induced modification of the coupled SWCNTs- Xe vibrational spectrum. The indicated change of the coefficient $C_{5}$ cannot readily be interpreted in terms of the bending rigidity without a model of the coupled SWCNT–adsorbate vibrational density of states. Nevertheless, changes in its value are consistent with the possibility that the rigidity of the SWCNTs is being modified.

This approach agrees with the recent work Barabashko \textit{et\,al.} \cite{barabashko2025experimental}, where it was shown that the heat capacity of carbon nanotube systems (SWCNT bundles, MWCNTs) at the lowest temperatures is systematically governed by two principal contributions: a Debye term, $C_{3}T^{3}$, arising from acoustic phonons, and a dispersive term, $C_{5}T^{5}$, associated with higher-order excitations~\cite{barabashko2025experimental}. Importantly, the negative value of the extracted $C_{5}$ coefficient provides evidence for flexural dispersion associated with out-of-plane ZA phonons,  agreement with theoretical predictions for low-dimensional carbon systems~\cite{kosevich2006crystal,gospodarev2017phonon,taheri2021importance}.

\begin{table}[htbp]
\centering
\begin{minipage}{0.85\textwidth}
\centering
\caption{Fitting parameters of the heat capacity $C(T)$ for pristine SWCNTs 
and SWCNT--gas composites. Listed are the coefficients $C_{3}$ and $C_{5}$, 
the adsorbed gas amount in cc\,STP, the  ratio 
$100\lvert C_{5}\rvert/C_{3}$, and the corresponding references for initial experimental data for reconstruction of heat capacity.}
\label{tab:fit_Cp}
\begin{tabular}{lcccccc}
\hline
Material & $C_{3}$ (mJ/g\,K$^{4}$) & $C_{5}$ (mJ/g\,K$^{6}$) &
Gas (cc STP) & $100\lvert C_{5}\rvert/C_{3}$ (K$^{-2}$) & Reference \\
\hline
SWCNT & $19\times10^{-3}$ & $-0.65\times10^{-3}$ & 0 & 3.42 & \cite{bagatskii2012specific} \\
SWCNT+Xe (7.15) & $47\times10^{-3}$ & $-1.4\times10^{-3}$ & 7.15 & 2.98 & \cite{bagatskii2013experimental}  \\
SWCNT+Ne (1.99) & $55\times10^{-3}$ & $-2.75\times10^{-3}$ & 1.99 & 5.00 & \cite{ramachandran2007thermodynamic} \\
SWCNT+Ne (3.98) & $75\times10^{-3}$ & $-3.32\times10^{-3}$ & 3.98 & 4.43 & \cite{ramachandran2007thermodynamic} \\
SWCNT+Ne (5.98) & $90\times10^{-3}$ & $-3.35\times10^{-3}$ & 5.98 & 3.72 & \cite{ramachandran2007thermodynamic} \\
SWCNT+Ne (11.91) & $105\times10^{-3}$ & $-3.45\times10^{-3}$ & 11.91 & 3.29 & \cite{ramachandran2007thermodynamic} \\
SWCNT+Ne (23.8) & $135\times10^{-3}$ & $-3.55\times10^{-3}$ & 23.8 & 2.63 & \cite{ramachandran2007thermodynamic} \\
SWCNT+Ne (29.7) & $167\times10^{-3}$ & $-3.62\times10^{-3}$ & 29.7 & 2.17 & \cite{ramachandran2007thermodynamic} \\
SWCNT+Ne (40.6) & $189\times10^{-3}$ & $-3.66\times10^{-3}$ & 40.6 & 1.94 & \cite{ramachandran2007thermodynamic} \\
SWCNT+Ne (46.6) & $223\times10^{-3}$ & $-5.49\times10^{-3}$ & 46.6 & 2.46 & \cite{ramachandran2007thermodynamic} \\
SWCNT+Ne (57.3) & $245\times10^{-3}$ & $-5.90\times10^{-3}$ & 57.3 & 2.41 & \cite{ramachandran2007thermodynamic} \\
SWCNT+Ne (74.4) & $257\times10^{-3}$ & $-5.70\times10^{-3}$ & 74.4 & 2.22 & \cite{ramachandran2007thermodynamic} \\
SWCNT+Ne (123.9) & $317\times10^{-3}$ & $-5.80\times10^{-3}$ & 123.9 & 1.83 & \cite{ramachandran2007thermodynamic} \\
SWCNT+Ne (153.6) & $338\times10^{-3}$ & $-5.95\times10^{-3}$ & 153.6 & 1.76 & \cite{ramachandran2007thermodynamic} \\
SWCNT+N$_2$ (8.54) & $35\times10^{-3}$ & $-0.85\times10^{-3}$ & 8.54 & 2.43 & \cite{bagatskii2013heat} \\
SWCNT+CH$_4$ (4.53) & -- & -- & 4.53 & -- & \cite{bagatskii2016heat} \\
\hline
\end{tabular}
\end{minipage}
\end{table}

Xenon adsorption produces a systematic increase in the Debye coefficient $C_{3}$ 
together with a pronounced rise in the absolute magnitude of the negative $C_{5}$ term, 
indicating a coupled modification of both acoustic and flexural phonon branches in 
SWCNT bundles. The enhancement of $C_{3}$ reflects the increased contribution of 
low-energy acoustic modes, which is consistent with Xe-induced mass loading of the 
nanotube framework and the associated changes of long-wavelength LA and TA phonons. 
Because the Debye term scales with the density of acoustic states near zero energy, 
even weak adsorbate–substrate coupling can alter the effective elastic modulus of the 
bundle and thereby increase the fitted value of $C_{3}$.

The large absolute value of negative $C_{5}$ coefficient observed for Xe-loaded bundles is 
attributed to Xe mass loading that impact in the total modification of the curvature of the ZA branch. It arises from the sensitivity of flexural 
(ZA-type) modes to localized perturbations of tube-tube and tube-Xe coupling. Xe atoms occupying 
groove and outer-surface sites introduce additional static mass and weak pinning, 
which modifies the curvature of the low-energy flexural dispersion and enhances the 
contribution of the ZA branch to the $T^{5}$ term in the heat capacity. This effect is 
particularly pronounced because flexural modes in SWCNT bundles possess intrinsically 
low stiffness and respond strongly to even modest symmetry breaking or local 
constraints. This interpretation is consistent with the experimental study of the radial thermal expansion \cite{dolbin2009radial}:  Xe impurity suppresses the negative contribution to the thermal expansion from the
ZA acoustic vibrations perpendicular to the nanotube surfaces.

The comparison between the heat capacity $C(T)$ and the radial thermal expansion 
$\alpha(T)$ \cite{dolbin2009radial} becomes fully consistent once the upward curvature of the flexural (ZA) 
phonon branch is taken into account. In SWCNT bundles, intertube coupling and 
residual tension introduce a linear long-wavelength component into the ZA dispersion, 
producing the characteristic upward curvature that naturally yields a negative $C_{5}$ 
coefficient in the low-temperature heat capacity. Xenon adsorption modifies this 
curvature through mass loading and weak pinning, thereby increasing the density of 
low-energy flexural states and enhancing the magnitude of the negative $C_{5}$ term. 
At the same time, the same adsorbate-induced pinning suppresses the anharmonic 
response of ZA vibrations, reducing their negative Grüneisen contribution \cite{schelling2003thermal} to the 
radial thermal expansion, in agreement with experimental observations. Thus, the 
opposite trends in $C(T)$ and $\alpha(T)$ \cite{dolbin2009radial} reflect the fact that heat capacity is governed 
by the phonon density of states, whereas thermal expansion is controlled by 
anharmonicity; Xe adsorption affects these two aspects of the ZA branch differently 
but in a mutually consistent manner.

As shown in Table~\ref{tab:fit_Cp}, Xe-loaded bundles exhibit the variation in the effective $C_{5}/C_{3}$ ratio in compare with nanotubes without adsorbates. Within the present phenomenological framework, the coefficients $C_{3}$ and $C_{5}$ characterize the low-temperature heat capacity of the coupled SWCNT--Xe system rather than individual phonon branches; therefore, changes in the $C_{5}/C_{3}$ ratio should be regarded as signatures of a modification of the low-energy excitation spectrum of the coupled system. A possible microscopic interpretation may be associated with the collective dynamics of Xe chains adsorbed in nanotube groove sites. According to the lattice-dynamical model proposed by Manzhelii \textit{et al.} \cite{manzhelii2025localized,manzhelii2019phonon}, adsorbed one-dimensional chains possess low-energy collective modes whose existence is intrinsically linked to their interaction with the nanotube bundle. Owing to the large mass of Xe atoms, the characteristic frequencies of these modes may approach the energy range of low-frequency transverse vibrations of SWCNT bundles. This raises the possibility of a partial hybridization between adsorbate-chain excitations and ZA-like modes of the nanotube framework. Such hybridization could renormalize the dispersion and spectral weight of the nanotube transverse modes, leading to changes in their contribution to the heat capacity and, consequently, to the observed evolution of the effective $C_{5}/C_{3}$ ratio. Within this scenario, the unusually large negative value of $C_{5}$ observed for Xe-loaded bundles may be viewed as indirect evidence for an adsorbate-induced restructuring of the low-energy vibrational spectrum, potentially involving modifications of ZA-like excitations through coupling with Xe-chain modes. However, this interpretation remains a physically motivated hypothesis consistent with the available calorimetric and thermal-expansion data and requires further verification by microscopic calculations and spectroscopic experiments.

Thus, the concurrent increase in $C_{3}$ and $|C_{5}|$ provides a physically consistent 
picture of adsorbate-induced changes in the phonon spectrum: acoustic modes are 
softened through mass loading, while flexural modes experience enhanced curvature due 
to weak pinning and modified intertube interactions. Together, these mechanisms 
account for the observed low-temperature heat-capacity behavior of Xe-decorated 
SWCNT bundles.

\subsection{Low temperature heat capacity of SWCNT bundles with adsorbed Ne }\label{subsec2}

\begin{figure}[htbp]
  \centering
  \includegraphics[width=0.85\textwidth]{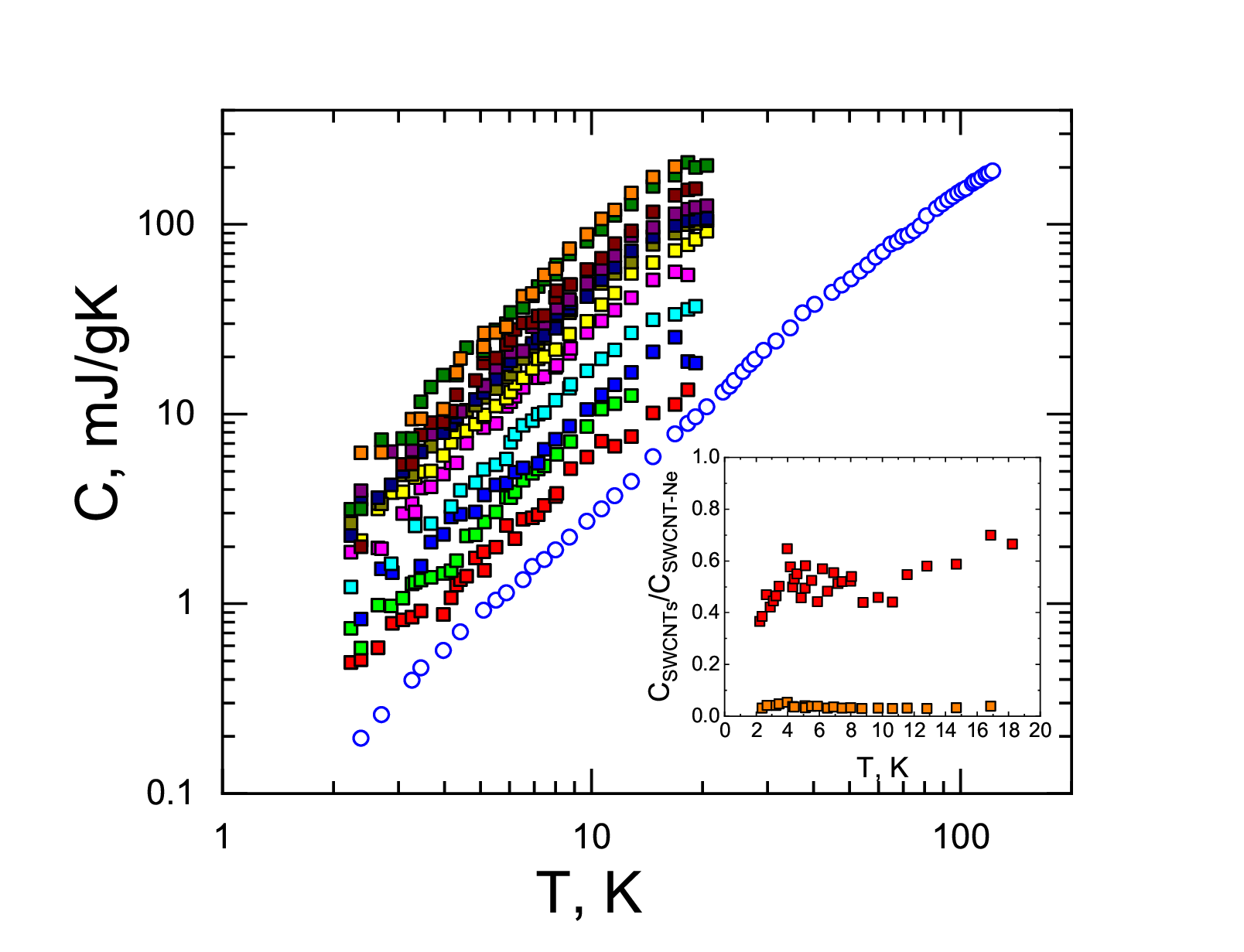}
  \caption{Temperature dependence of the heat capacity of pristine c-SWCNT bundles (\textcolor{blue}{\(\circ\)}), and c-SWCNT bundles with adsorbed Ne at various Ne loadings: \textcolor{neRed}{\(\blacksquare\)}\,1.99, \textcolor{neGreen}{\(\blacksquare\)}\,3.98, \textcolor{neBlue}{\(\blacksquare\)}\,5.98, \textcolor{neCyan}{\(\blacksquare\)}\,11.91, \textcolor{neMagenta}{\(\blacksquare\)}\,23.8, \textcolor{neYellow}{\(\blacksquare\)}\,29.7, \textcolor{neDarkYellow}{\(\blacksquare\)}\,40.6, \textcolor{neNavy}{\(\blacksquare\)}\,46.6, \textcolor{nePurple}{\(\blacksquare\)}\,57.3, \textcolor{neWine}{\(\blacksquare\)}\,74.4, \textcolor{neOlive}{\(\blacksquare\)}\,123.9, \textcolor{neOrange}{\(\blacksquare\)}\,153.6\,scc. Increasing Ne concentration produces a systematic rise in the total heat capacity and in the $C_{3}$ and $\lvert C_{5} \rvert$ terms across the entire temperature range (see Table~\ref{tab:fit_Cp}). The inset shows the ratio $C_{\mathrm{SWCNT}}/C_{\mathrm{SWCNT\text{-}Ne}}$ for the lowest (\textcolor{neRed}{\(\blacksquare\)}\,1.99\,scc) and highest (\textcolor{neOrange}{\(\blacksquare\)}\,153.6\,scc) Ne coverages, illustrating the evolution of the nanotube contribution to the heat capacity of the system and the disappearance of the characteristic low-temperature kink at high adsorbate density.}

  \label{fig:fig_4}
\end{figure}

Figure~\ref{fig:fig_4} presents the total heat capacity of the SWCNT--Ne system calculated using the mixture rule (Eq.~~\ref{eq:mixture}), combining the heat capacity of pristine c-SWCNT bundles~\cite{bagatskii2012specific} with the additive contributions of Ne adsorbate films measured by Ramachandran and Vilches~\cite{ramachandran2007thermodynamic} for coverages ranging from 1.99\,scc to 153.6\,scc. The resulting heat capacity of the system reflects modifications of the Ne longitudinal and transverse vibrational modes, alterations of the SWCNT acoustic and flexural branches, and enhanced coupling between the adsorbate and the nanotube lattice. For comparison, the heat capacities of pristine c-SWCNT bundles is also shown. In all cases (SWCNTs~\cite{bagatskii2012specific} and SWCNT--Ne~\cite{ramachandran2007thermodynamic}), the mean nanotube diameter was approximately 1.1\,nm.

Even the smallest Ne concentration produces a measurable increase in the total heat capacity across the entire temperature range, whereas higher Ne loadings lead to a pronounced and systematic enhancement of the heat capacity of the system. The inset of Fig.~~\ref{fig:fig_4} displays the ratio $C_{\mathrm{SWCNT}}/C_{\mathrm{SWCNT+gas}}$, highlighting the relative contribution of the nanotube matrix. For the lowest Ne coverage (1.99\,scc), this ratio remains close to 50\,\% and exhibits a distinct kink (change in slope)  near 10\,K, similar to the feature observed for the SWCNT--Xe system. This kink indicates that, at low adsorbate densities, Ne atoms occupy groove sites and other strongly binding outer-surface positions, where their dynamics are partially suppressed and the adsorbate behaves predominantly as a static mass load at the lowest temperatures. The appearance of the kink  can indicate coupling between the adsorbate particles and the flexural (ZA) phonon branch of the SWCNT lattice, whose out-of-plane displacements are highly sensitive to perturbations introduced by Ne atoms at the nanotube surface.

For a loading of 3.98\,scc of Ne, the experimental heat capacity of the adsorbate subsystem agrees, in the 2--4\,K interval, with the calculated heat capacity of an ideal one-dimensional (1D) Ne chain~\cite{kostov2003phonons}, enabling an estimate of the Debye temperature associated with 1D vibrational modes. Above 4\,K, however, the discrepancy between experiment and theory increases systematically, suggesting that the real adsorbate environment deviates from the idealized 1D model. A pronounced difference exists between the experimentally derived Debye temperature, $\Theta_{0}=58$\,K~\cite{ramachandran2007thermodynamic}, and the theoretical values, which span $\Theta_{0}\approx 78$--120\,K depending on the assumed adsorbate--adsorbate force constants and ground-state spacing (3.08\,\AA{} or 2.92\,\AA{} for Ne) in the modeled 1D lattice~\cite{kostov2003phonons}. Although the theoretical $\Theta_{0}$ is highly sensitive to these parameters, this sensitivity alone may not account for the markedly lower experimental value. A plausible origin of the deviation is that the idealized 1D model assumes adsorption exclusively in groove (G) sites, whereas in real SWCNT bundles neon atoms---similarly to helium---occupy both groove sites and interstitial channels (IC) between adjacent nanotubes. These IC sites, which are not included in the theoretical model, possess higher binding energies and transverse dimensions larger than the effective diameters of light adsorbates. As shown by Stan and co-workers~\cite{stan1998interstitial,stan2000uptake}, light impurities such as $^{4}$He, H$_2$, and Ne preferentially fill IC sites at low temperatures. Occupation of these additional adsorption sites can modify both the vibrational spectrum of the Ne adsorbate and the phonon spectrum of the SWCNTs lattice by altering van der Waals interactions, intertube coupling, and the effective bending rigidity of the bundle.

In contrast, at high Ne coverages (153.6\,scc) approaching one monolayer, the ratio $C_{\mathrm{SWCNT}}/C_{\mathrm{SWCNT+gas}}$ decreases to only $\sim 5\%$, and the kink near 10\,K disappears entirely. When the nanotube surface becomes densely populated, Ne atoms form a quasi-continuous adsorbate layer with a substantially larger vibrational density of states. In this regime, the adsorbate subsystem dominates the total heat capacity throughout the entire temperature interval, and the characteristic signature of ZA-mode--adsorbate coupling is effectively masked. The absence of the kink at high coverage indicates that the flexural phonon contribution of the SWCNT lattice becomes comparatively small within the total heat capacity of the SWCNT--Ne system, while the vibrational modes of the Ne adsorbate layer overwhelm the nanotube contribution.

The evolution of this ratio with Ne loading suggests that adsorption progressively changes the balance between host and adsorbate contributions. At low coverage, the adsorbed atoms constitute a relatively small perturbation and the low-temperature heat capacity retains a substantial contribution from the nanotube framework. At high coverage, the large number of adsorbed atoms introduces a correspondingly large density of low-energy excitations, and the intrinsic contribution of the nanotube lattice becomes comparatively small. The disappearance of the feature near 10 K at the highest loading should therefore not be interpreted solely as the disappearance of a particular SWCNTs phonon branch. Rather, it is consistent with the characteristic features associated with the nanotube contribution becoming progressively masked by the much larger heat capacity of the adsorbate-containing system.

\begin{figure}[htbp]
  \centering

  \begin{subfigure}{0.75\textwidth}
    \centering
    \includegraphics[width=\linewidth]{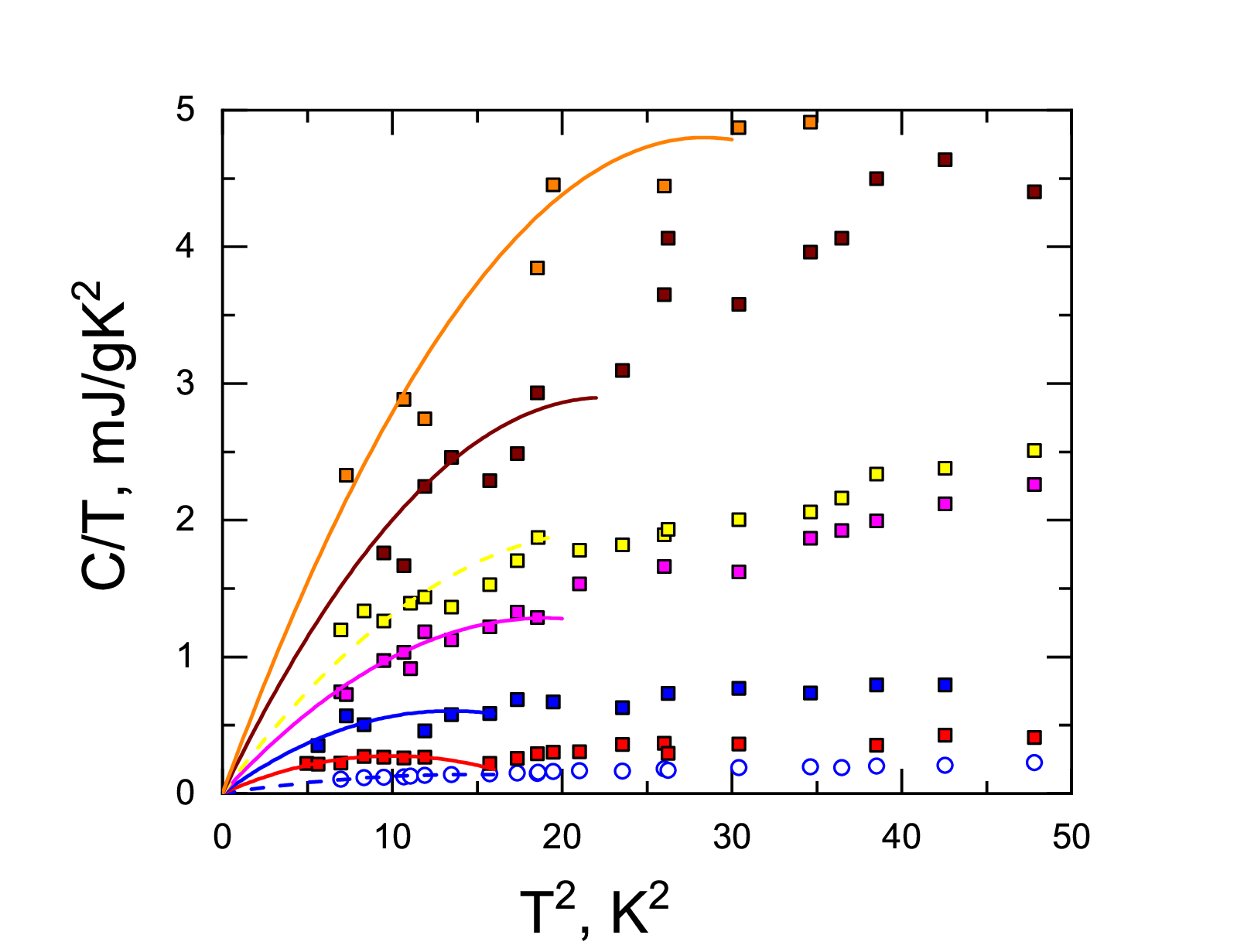}
    \caption{$C/T$ plotted as a function of $T^{2}$.}
    \label{fig:fig_5a}
  \end{subfigure}

  \vspace{0.5\baselineskip}

  \begin{subfigure}{0.75\textwidth}
    \centering
    \includegraphics[width=\linewidth]{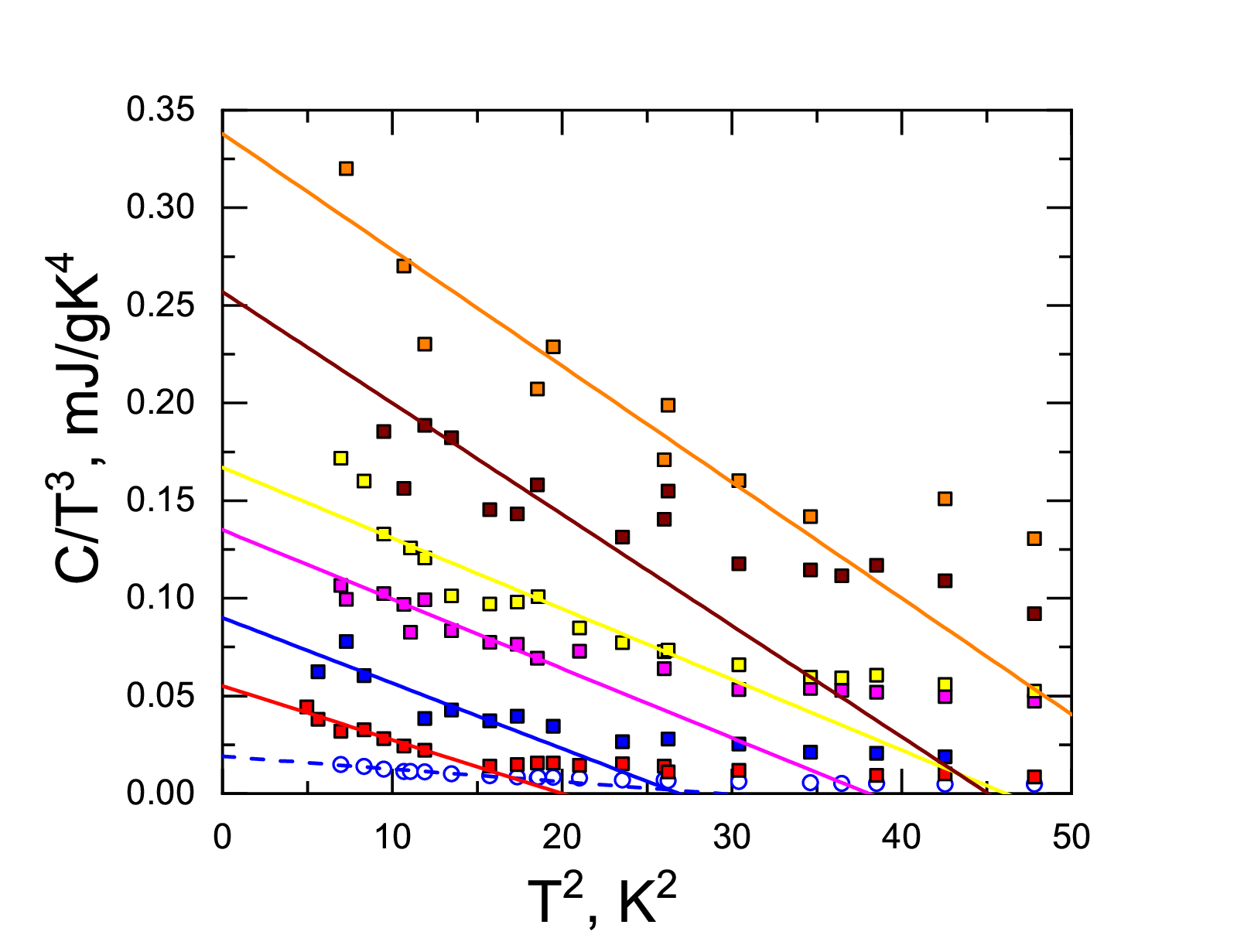}
    \caption{$C/T^{3}$ plotted as a function of $T^{2}$.}
    \label{fig:fig_5b}
  \end{subfigure}

  \caption{Temperature dependence of (a) $C/T$ versus $T^{2}$ and (b) $C/T^{3}$ versus $T^{2}$ for pristine c-SWCNT bundles (\textcolor{blue}{\(\circ\)}) and selected c-SWCNT bundles with adsorbed Ne at representative loadings: \textcolor{neRed}{\(\blacksquare\)}\,1.99\,scc, \textcolor{neBlue}{\(\blacksquare\)}\,5.98\,scc, \textcolor{neMagenta}{\(\blacksquare\)}\,23.8\,scc, \textcolor{neYellow}{\(\blacksquare\)}\,29.7\,scc, \textcolor{neWine}{\(\blacksquare\)}\,74.4, \textcolor{neOrange}{\(\blacksquare\)}\,153.6\,scc. The solid lines represent polynomial fits to the experimental data using Eq.~\eqref{eq:CT} for the pristine SWCNT and SWCNT--Ne systems, with the corresponding $C_{3}$ and $C_{5}$ parameters listed in Table~\ref{tab:fit_Cp}. The remaining Ne loadings were analyzed analogously, and the fitted coefficients $C_{3}$ and $C_{5}$ are summarized in Table~\ref{tab:fit_Cp}.}
  \label{fig:fig_5}
\end{figure}

As for the pristine SWCNT and SWCNT--Xe systems, the Ne-containing composites were analyzed below approximately 7\,K using the $C/T$ versus $T^{2}$ representation (Fig.~~\ref{fig:fig_5}a) and the
complementary $C/T^{3}$ versus $T^{2}$ representation (Fig.~~\ref{fig:fig_5}b). Below approximately 4\,K,
the data for the different Ne loadings are described to good accuracy by
\begin{equation}
C(T) = C_{3}T^{3} + C_{5}T^{5}.
\label{eq:CT}
\end{equation}
The mutually consistent behavior in the two representations supports the view that this two-term
expansion provides a useful phenomenological description of the low-energy heat capacity of
the SWCNT--Ne system. $C_{5}^{(0)}$ is an effective dispersive coefficient, associated in pristine
SWCNT bundles with the low-energy flexural ZA phonons. $C_{3}^{(0)}$ captures the Debye-like acoustic contribution to the heat capacity of pristine SWCNT bundles. The effective coefficients of the coupled SWCNT--Ne system ($C_{3}$ and $C_{5}$) are summarized in
Table~\ref{tab:fit_Cp}. It can be seen that Ne loading systematically decreases the ratio $C_{5}/C_{3}$, indicating a change in their effective balance.

A systematic evolution of both coefficients is observed with increasing Ne loading (Table~\ref{tab:fit_Cp}). The
effective $C_{3}$ coefficient increases as the amount of adsorbed Ne increases, indicating an
enhanced low-energy density of states relative to the pristine SWCNT reference. In a
conventional Debye description, $C_{3}$ is related to the characteristic acoustic velocities.
However, the present SWCNT--Ne system is structurally heterogeneous and contains both
nanotube and adsorbate degrees of freedom; consequently, the observed increase in $C_{3}$
should be interpreted as an effective change in the combined low-energy acoustic response
rather than as a direct measurement of a single elastic constant of the nanotube lattice.

The evolution of $C_{5}$ is particularly significant. The negative $C_{5}$ coefficient becomes
progressively larger in magnitude with increasing Ne loading. Since the pristine SWCNT bundles
already exhibit a negative $C_{5}$, the observed concentration dependence indicates that the
dispersive component of the low-energy heat capacity is systematically modified by adsorption.
This result is difficult to reconcile with a strictly additive model in which the pristine SWCNT
contribution remains invariant and the Ne subsystem simply contributes an independent heat
capacity.

Several microscopic effects may contribute to this evolution. Ne adsorption changes the mass
distribution in the vicinity of the nanotube surfaces and introduces additional vibrational
degrees of freedom. At the same time, occupation of groove and interstitial sites can modify
the local van der Waals environment and the mechanical coupling between neighboring
nanotubes. These perturbations can influence the low-frequency acoustic and flexural modes of
the bundle. In addition, vibrational modes localized predominantly on the adsorbate may
interact with the low-energy modes of the nanotube framework. The experimentally determined
$C_{5}$ should therefore be regarded as an effective coefficient reflecting the combined
dispersive response of the coupled SWCNT--Ne system.

These observations are consistent with the temperature dependence of the thermal expansion of SWCNT bundles saturated with gases~\cite{dolbin2009radial,dolbin2010low}. Gas adsorption suppresses the amplitude of out-of-plane (ZA) vibrations by increasing the effective mass density of the nanotube wall and stiffening the nanotube surface. At low Ne coverage, the ZA mode remains well defined, and its quadratic dispersion produces a noticeable kink in the heat-capacity ratio. As coverage increases, the adsorbate layer modifies the boundary conditions for out-of-plane vibrations and progressively changes the effective balance between the flexural ZA phonons and LA, TA branches of SWCNTs. The reduction of 
$C_{\mathrm{SWCNT}}/C_{\mathrm{SWCNT+gas}}$
and the disappearance of the kink therefore reflect the gradual suppression of the ZA branch by the increasingly loading of the Ne layer.

Thermal-expansion measurements by Dolbin \textit{et al.}~\cite{dolbin2009radial,dolbin2010low,vinnikov2023hydrogen} further indicate that gas adsorption can significantly alter the low-energy vibrational dynamics of SWCNT bundles. The formation of bundles and the adsorption of impurity atoms on external surfaces, in groove sites, or inside nanotubes introduce additional intertube and adsorbate--nanotube interactions, thereby enhancing the three-dimensional character of the system. As a result, the negative Gr\"uneisen parameters commonly associated with low-frequency transverse vibrations ~\cite{schelling2003thermal} decrease in magnitude or may even become positive. Consequently, the negative contribution to the radial thermal expansion of Xe- and N$_2$-loaded bundles is reduced and shifted toward lower temperatures~~\cite{dolbin2010low}. Combined with the calorimetric results presented here, these observations suggest an adsorbate-induced restructuring of the low-energy excitation spectrum. In the particular case of Xe, the large adsorbate mass and the existence of low-frequency collective modes in adsorbed Xe chains~\cite{manzhelii2019phonon} raise the possibility of a renormalization and partial hybridization of ZA-like nanotube excitations with adsorbate-chain modes. While this interpretation remains hypothetical, it provides a physically plausible framework for understanding the correlated changes observed in both heat-capacity and thermal-expansion measurements.

The heat-capacity and thermal-expansion results therefore point in the same general direction:
gas adsorption does not simply introduce an independent set of additional degrees of freedom,
but perturbs the low-energy vibrational response of the nanotube bundle. The present
heat-capacity analysis provides a quantitative phenomenological measure of this effect through
the evolution of $C_{3}$ and $C_{5}$, while the thermal-expansion data provide an independent indication of changes in the low-energy out-of-plane ZA transverse modes.

An important feature of the Ne data is that the $C_{3}T^{3}+C_{5}T^{5}$ representation remains
applicable despite the very large variation in adsorbate concentration. At low Ne loadings, the
expansion provides an adequate description only up to approximately 4\,K. Above this
temperature, contributions from higher-energy vibrational states of the nanotube bundle become
increasingly important, and systematic deviations from Eq.~\ref{eq:CT}) appear. At the highest Ne
loading, approximately 153.6\,scc, the same expansion remains applicable up to about 5.5\,K.
This extension of the fitting interval does not necessarily imply that the intrinsic SWCNTs phonon
spectrum remains harmonic over a wider range. Rather, it reflects the fact that the dominant
low-energy excitations of the heavily loaded composite have been substantially reorganized by
the presence of the Ne layer, while the relative contribution of the pristine-SWCNT
higher-energy modes to the total heat capacity is reduced.

Thus, the persistence of the $C_{3}T^{3}+C_{5}T^{5}$ form over a broad range of Ne concentrations
is best regarded as a robust empirical property of the low-energy heat capacity of the coupled
SWCNT--Ne system. The coefficients themselves, however, are not universal: both $C_{3}$ and
$C_{5}$ evolve systematically with Ne loading. This distinction is essential. The stability of the
functional form does not imply invariance of the underlying phonon spectrum; rather, the
systematic evolution of its coefficients suggests that the spectrum represented by this
effective expansion changes continuously with adsorption.

The Ne series consequently provides relatively strong evidence for adsorption-induced modification of
the low-energy vibrational response compared with a comparison of only two samples. The
continuous increase of the effective $C_{3}$ and the magnitude of the negative $C_{5}$ with Ne
loading indicates a pronounced concentration-dependent trend. While the present calorimetric data
do not uniquely identify the microscopic origin of this trend, they suggest that the
low-energy heat capacity of the SWCNT bundle evolves systematically with adsorption and
may not, over the entire concentration range, be interpreted simply as the heat capacity of an
unchanged pristine SWCNTs host plus an independent Ne contribution.

\subsection{Low-temperature heat capacity of SWCNT bundles with molecular adsorbates: N$_2$ and CH$_4$}
\label{subsec3}

\begin{figure}[htbp]
  \centering
  \includegraphics[width=0.85\textwidth]{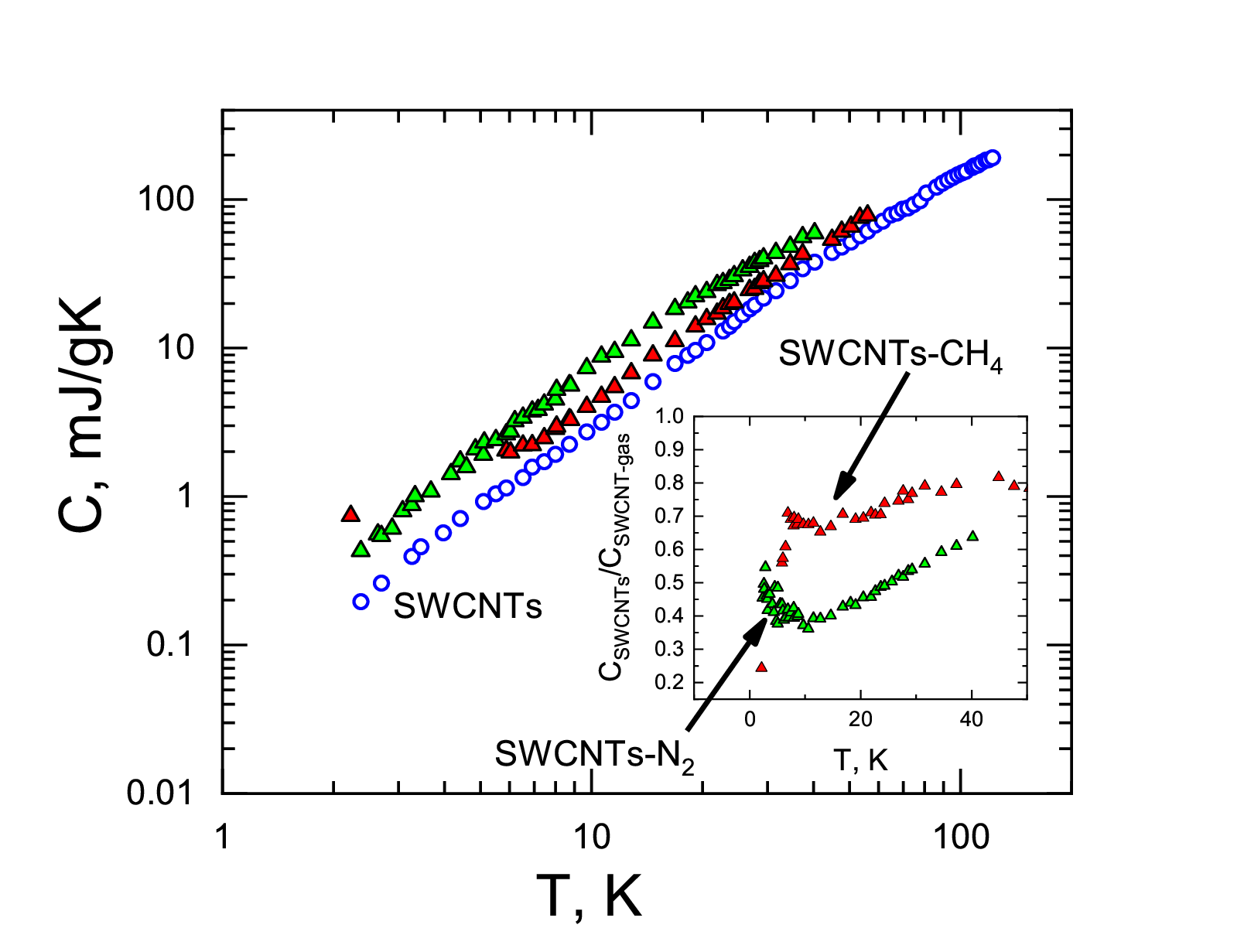}
  \caption{Temperature dependence of the heat capacity of pristine c-SWCNT bundles (\textcolor{blue}{$\circ$}), SWCNT bundles with adsorbed N$_2$ (\textcolor{green}{$\blacktriangle$}), and SWCNT bundles with adsorbed CH$_4$ (\textcolor{red}{$\blacktriangle$}). The data are shown on double-logarithmic axes to highlight the differences in the low-temperature behavior of the three systems. The inset shows the ratio $C_{\mathrm{SWCNT}}/C_{\mathrm{SWCNT\text{-}gas}}$ for the SWCNT bundles with adsorbed N$_2$ (\textcolor{green}{$\blacktriangle$}), and SWCNT bundles with adsorbed CH$_4$ (\textcolor{red}{$\blacktriangle$}). }
 \label{fig:fig_6}
\end{figure}

Figure~\ref{fig:fig_6} shows the temperature dependence of the heat capacity of pristine closed-end
SWCNT bundles with that of SWCNT bundles containing adsorbed N$_2$ and CH$_4$. The data are
presented on double-logarithmic axes, while the inset shows the ratio
$C_{\mathrm{SWCNT}}/C_{\mathrm{SWCNT+gas}}$. This representation provides a direct comparison
of the relative contributions of the nanotube host and the adsorbed species and illustrates how
the thermodynamic response evolves with temperature and adsorbate character.

In all three systems, adsorption produces a substantial enhancement of the total heat capacity relative to pristine SWCNT bundles, particularly at low temperatures. This enhancement may not, in general, be interpreted as a simple universal additive correction because the adsorbates
differ qualitatively in their internal degrees of freedom and in their coupling to the nanotube
substrate. N$_2$ and CH$_4$ are molecular species with translational, rotational, and
orientational degrees of freedom \cite{bagatskii1968thermal, minchina2001heat, bagatskii2003rotational,krivchikov1988dilute, bagatskii2014low}. These additional molecular degrees of freedom can make an
appreciable contribution to the heat capacity and lead to a substantially different temperature
dependence of the SWCNT--gas systems \cite{bagatskii2016heat,bagatskii2017heat}. The heat capacity of the
SWCNT--CH$_4$ system lies substantially above that of pristine SWCNTs over a broad temperature
interval. This behavior is consistent with the known thermodynamics of confined methane, for
which rotational and orientational degrees of freedom contribute strongly to the heat capacity.
In one-dimensional or strongly confined CH$_4$ systems, the rotational subsystem may exhibit a
low-temperature Schottky-type anomaly, an orientational transition, and a crossover toward a
free-rotor regime at higher temperatures \cite{bagatskii2016heat,bagatskii2017heat}. These features suggest that the molecular
contribution may not be represented simply as an additional acoustic phonon term; rather, it
constitutes a distinct set of low-energy excitations superimposed on, and potentially coupled to,
the vibrational spectrum of the nanotube host.

The behavior of the N$_2$-containing system is somewhat different. At the lowest temperatures,
its heat capacity is dominated primarily by translational excitations, whereas orientational and
librational degrees of freedom become increasingly important at temperatures of approximately
10\,K~\cite{dolbin2009radial,dolbin2010low, bagatskii2013heat}. Consequently, the relative contribution of the nanotube
matrix decreases more gradually with increasing temperature than in the CH$_4$ system. The
different temperature dependences of the N$_2$ and CH$_4$ composites therefore provide an
important indication that the molecular character of the adsorbate, rather than its mass alone,
plays a significant role in determining the low-temperature thermodynamics.

The inset of Fig.~\ref{fig:fig_6} emphasizes this difference particularly well. For the CH$_4$-loaded
sample, the ratio $C_{\mathrm{SWCNT}}/C_{\mathrm{SWCNT+gas}}$ is strongly reduced, indicating
that the adsorbate makes a substantial contribution to the measured heat capacity. The
suppression is less pronounced for N$_2$, consistent with the weaker contribution of its
rotational degrees of freedom at the lowest temperatures. It is important to note that, as in the cases of Xe and Ne adsorption, a distinct change in the slope of the heat-capacity curve is observed near 10~K. Adsorption produces a pronounced enhancement of the heat capacity, and this low-temperature feature appears to be reproducible regardless of the adsorbate species. Below approximately 10~K, the adsorbed particles are strongly confined by the adsorption potential of the nanotube surface and therefore behave predominantly as localized mass loads coupled to the SWCNT lattice. In this regime, their dynamics are largely suppressed, and the observed $C_{\mathrm{SWCNT}}/C_{\mathrm{SWCNT+gas}}$ reflects the interaction between the immobilized adsorbate and the low-energy vibrational modes of the nanotube bundle. The increase in the distance between neighboring xenon atoms for chains adsorbed in the grooves of SWCNT bundles at temperatures above 8~K was discussed by Bagatskii et al.~\cite{bagatskii2013experimental}. This observation suggests that the
additional heat capacity is difficult to attribute exclusively to molecular rotations or librations in the case of SWCNT bundles with molecular adsorbates.
The presence of the adsorbate itself, together with
its mass loading and interaction with the nanotube bundle, can significantly influence the
low-energy thermodynamic response of SWCNT bundles.

An important point follows from this comparison. The additive representation introduced in
Eq.~\ref{eq:additive} provides a useful baseline, but it should not be interpreted as proof that the vibrational
spectra of the nanotube and adsorbate subsystems remain independent. The observed heat
capacity represents the response of the coupled SWCNT--gas system. Adsorption can alter the
effective mass distribution, intertube interactions, local force constants, and boundary
conditions for nanotube vibrations, while the adsorbate simultaneously introduces its own
translational, rotational, and librational excitations. The measured heat capacity may therefore
contain both an approximately additive adsorbate contribution and a non-additive modification
of the host vibrational spectrum.

\begin{figure}[htbp]
  \centering
  \includegraphics[width=0.75\textwidth]{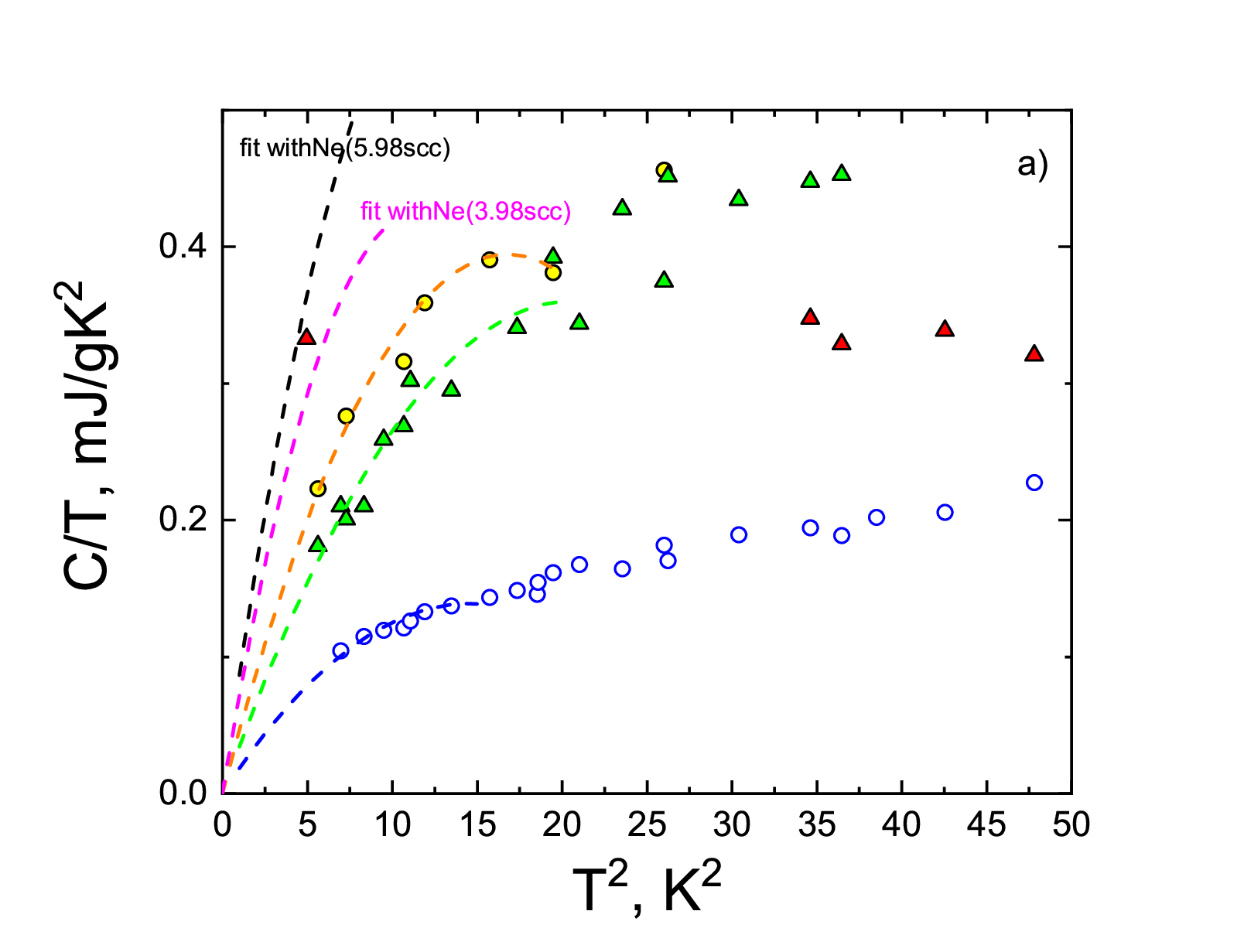}
  \includegraphics[width=0.75\textwidth]{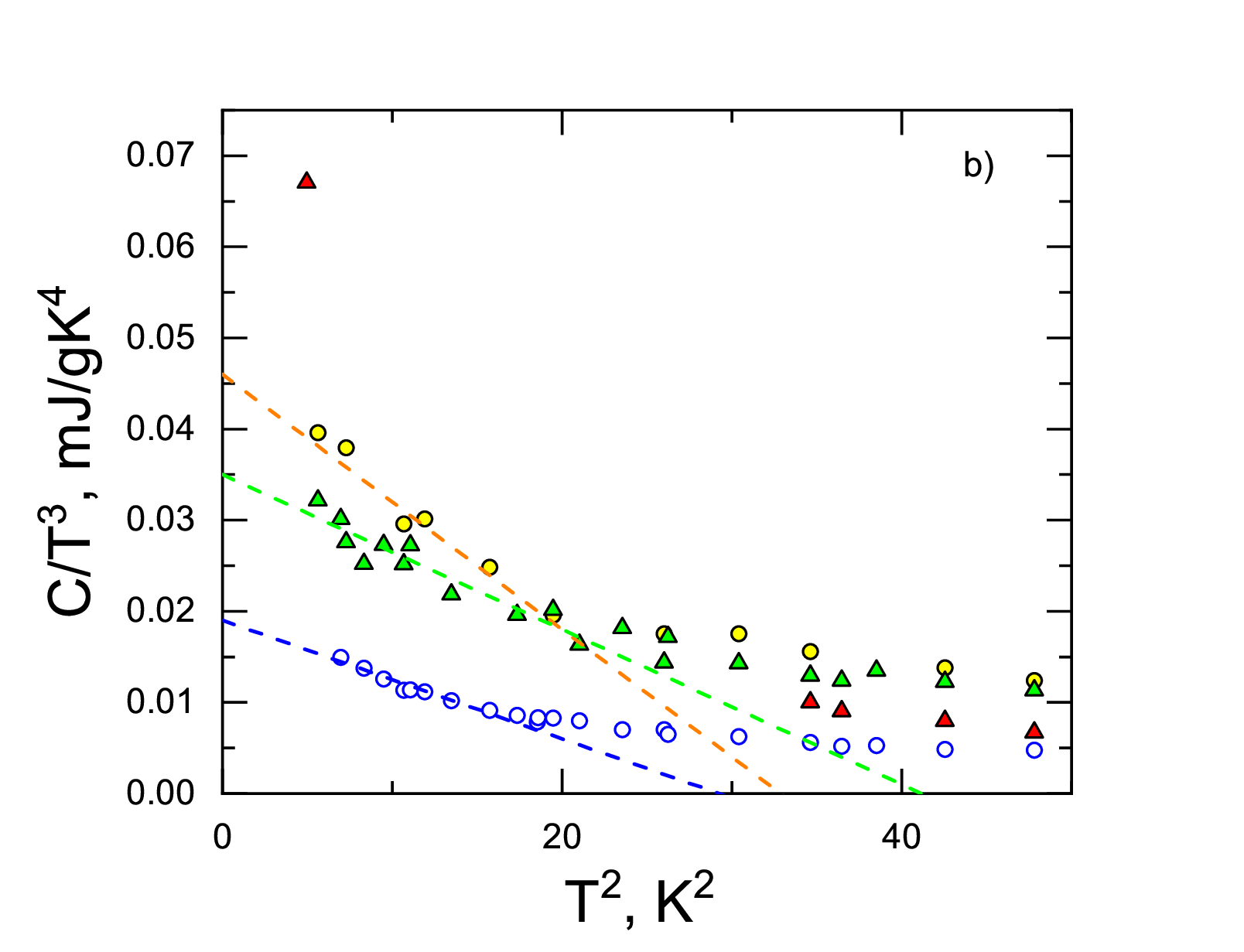}
  \caption{
  (a) Temperature dependence of $C/T$ plotted as a function of $T^{2}$, and (b) temperature
  dependence of $C/T^{3}$ versus $T^{2}$, for pristine c‑SWCNT bundles
  (\textcolor{blue}{\(\circ\)}), c‑SWCNT bundles with adsorbed Xe
  (\textcolor{yellow}{\(\bullet\)}), N$_2$ (\textcolor{green}{\(\blacktriangle\)}), and CH$_4$
  (\textcolor{red}{\(\blacktriangle\)}). The straight lines represent polynomial fits to the
  experimental data using Eq.~\eqref{eq:CT} for the SWCNT, SWCNT–Xe, SWCNT-N$_2$ and two concentrations of SWCNT–Ne (3.98 scc and 5.98 scc) with the fitted $C_{3}$ and $C_{5}$ parameters listed in Table~\ref{tab:fit_Cp}. The SWCNT–CH$_4$ (4.53 scc) has apparently the close parameters as to those of SWCNT–Ne (3.98 scc and 5.98 scc) systems. However, a reliable determination of the coefficients $C_{3}$ and $C_{5}$ for the SWCNT–CH$_4$ system is not possible because the available dataset contains an insufficient number of experimental points below 3 K for this system.}
  \label{fig:fig_7}
\end{figure}

The low-temperature heat capacities of the c-SWCNT–Xe, c-SWCNT–N$_2$, and c-SWCNT–CH$_4$
composites were analyzed below 7\,K using the complementary $C/T$ versus $T^{2}$ and
$C/T^{3}$ versus $T^{2}$ representations (Figs.~\ref{fig:fig_7}a and ~\ref{fig:fig_7}b), which provide a sensitive means of
extracting the effective coefficients $C_{3}$ and $C_{5}$ of the polynomial expansion in
Eq.~\eqref{eq:CT}. In the $C/T$ versus $T^{2}$ plots, (c-SWCNT–Xe, c-SWCNT–N$_2$) systems exhibit a well-defined linear region below
4--5\,K, supporting the dominance of the Debye $T^{3}$ term and enabling a reliable
determination of $C_{3}$. The slopes increase systematically from pristine SWCNTs to N$_2$-, Xe- 
loaded SWCNT bundles, reflecting the enhancement of the low-energy density of states
through mass loading, acoustic stiffening, and low-energy translational excitations of the
adsorbates. The complementary $C/T^{3}$ versus $T^{2}$ representation highlights the dispersive
$T^{5}$ contribution associated with flexural (ZA) phonons and low-energy adsorbate excitations:
adsorption shifts the curves upward, indicating systematic changes in the effective $C_{5}$
coefficient. For Xe, this modification arises primarily from mass loading and suppression of
intrinsic ZA modes, whereas for N$_2$ and CH$_4$ the effect is stronger due to additional
rotational and librational degrees of freedom that become thermally populated with increasing
temperature. CH$_4$ produces the largest deviation, consistent with its rich rotational dynamics \cite{bagatskii2016heat,bagatskii2017heat},
while N$_2$ shows an intermediate behavior with rotational contributions emerging above
10\,K \cite{bagatskii2013heat}. A reliable determination of the coefficients $C_{3}$ and $C_{5}$ for the SWCNT–CH$_4$ system is not possible due to the available dataset contains an insufficient number of experimental points below 3 K. 

These observations indicate that, despite the presence of molecular adsorbates with
additional internal degrees of freedom, the combined SWCNT--gas systems retain a well-defined
low-temperature polynomial form. The extracted coefficients should therefore be regarded as
effective parameters of the coupled system rather than intrinsic properties of the nanotube
lattice. In particular, changes in $C_{5}$ are consistent with modifications of the net dispersive term of the
phonon spectrum, which may include contributions from altered flexural modes, low-energy
molecular excitations, and their hybridization. 
Together, the results of Fig.~\ref{fig:fig_7} suggest that the same low-temperature phenomenology
remains applicable across all adsorbates, while its effective coefficients evolve systematically
with the nature and internal dynamics of the adsorbed species.

Consequently, the evolution of $C_{tot}(T)$ may be difficult to interpret within a strictly additive model. A more appropriate description is

\begin{equation}
C_{\mathrm{tot}}(T) =
C_{\mathrm{SWCNT}}(T) +
C_{\mathrm{ads}}(T) +
\Delta C_{\mathrm{int}}(T),
\label{eq:total_heat_capacity}
\end{equation}

where the additional term $\Delta C_{\mathrm{int}}(T)$ accounts for adsorption-induced
modifications of the host SWCNTs spectrum and nanotube--adsorbate coupling. The progression from Xe
to N$_2$ and CH$_4$ suggests a hierarchy of spectral reorganization: Xe primarily alters
the acoustic branches through mass loading, N$_2$ introduces orientational and librational
excitations, and CH$_4$ tends to produce the strongest modification due to its pronounced rotational
dynamics. 

\subsection{Discussion of adsorption-induced modification of the low-energy spectrum of SWCNTs-gas system}\label{subsec4}

\begin{figure}[htbp]
  \centering
  \includegraphics[width=0.85\textwidth]{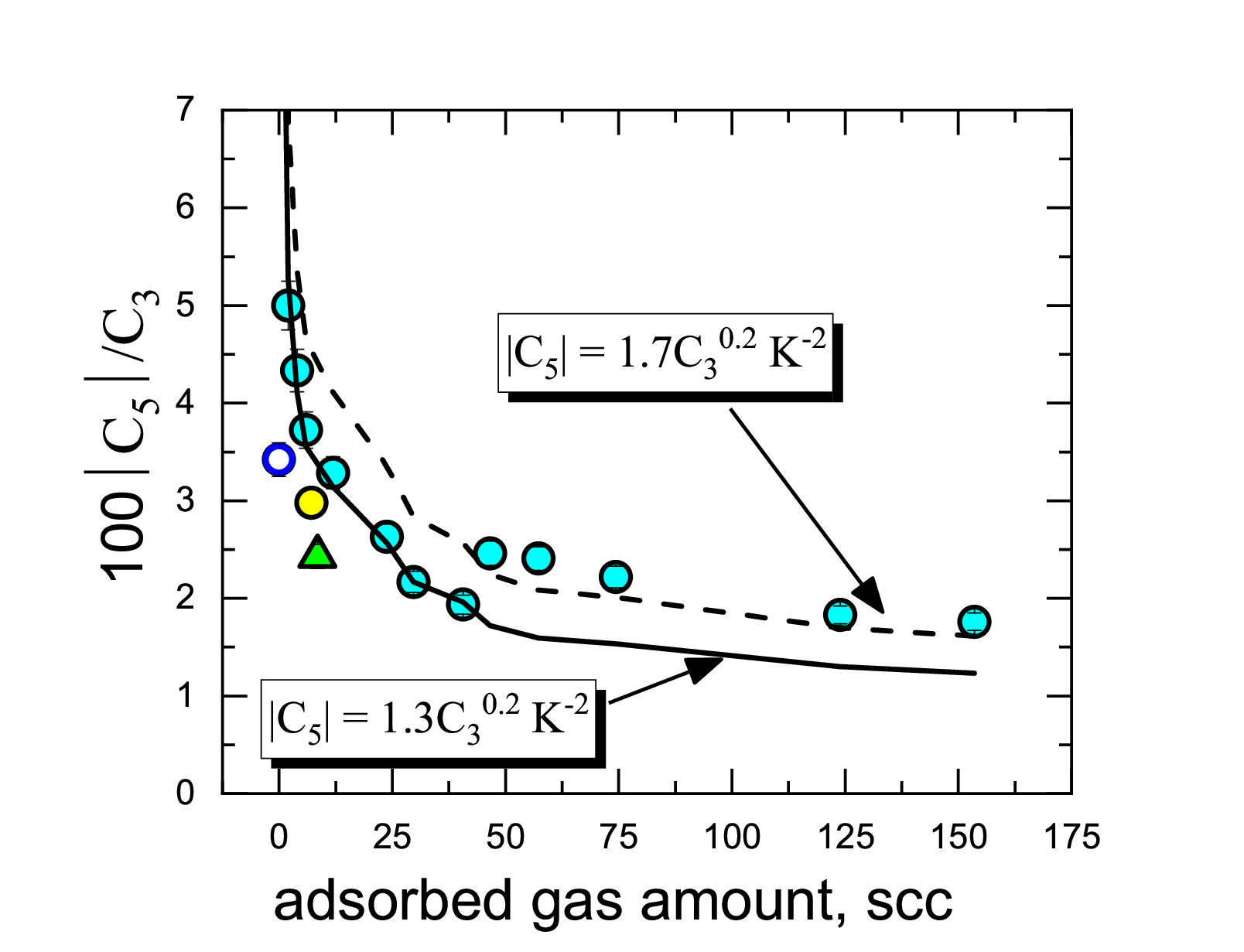}
  
  \caption{Dependence of the ratio $100|C_{5}|/C_{3}$ on the amount of adsorbed gas for the SWCNT--gas systems (SWCNT--Ne, SWCNT--Xe, and SWCNT--N$_2$), shown for pristine c-SWCNT bundles (\textcolor{blue}{\(\circ\)}), Xe-loaded SWCNT bundles (\textcolor{yellow}{\(\bullet\)}), Ne-loaded SWCNT bundles (\textcolor{cyan}{\(\bullet\)}), and N$_2$-loaded SWCNT bundles (\textcolor{green}{\(\blacktriangle\)}). The solid and dashed lines serve as guides to the eye and illustrate the systematic decrease of the ratio $100|C_{5}|/C_{3}$ with increasing adsorbate loading. The solid black line corresponds to the scaling $\lvert C_{5}\rvert = 1.3\times C_{3}^{0.2}\,\mathrm{K^{-2}}$ characteristic of SWCNT bundles with Ne loading. The dashed line corresponds to the scaling $\lvert C_{5}\rvert = 1.7\times C_{3}^{0.2}\,\mathrm{K^{-2}}$ characteristic of SWCNT bundles with Ne loading. The fitted $C_{3}$ and $C_{5}$ parameters are listed in Table~\ref{tab:fit_Cp}.}
  \label{fig:fig_8}
\end{figure}

Figure~\ref{fig:fig_8} shows the dependence of the ratio $100|C_{5}|/C_{3}$ on the amount of adsorbed gas for the SWCNT--gas systems (SWCNTs--Ne, SWCNTs--Xe, and SWCNTs--N$_2$). A pronounced
decrease trend of $100|C_{5}|/C_{3}$ is observed as the adsorbate concentration increases
from a few scc to more than 150\,scc, and this trend is well captured by the dashed guide-to-the-eye line, which highlights the systematic suppression of the dispersive $C_{5}$ term relative to the Debye $C_{3}$ term with increasing loading.

At low coverages, isolated adsorbates act as localized perturbations and the flexural ZA branch
remains well defined, yielding a relatively large dispersive contribution. As the coverage increases,
mass loading, surface pinning, and stiffening of the nanotube walls progressively suppress
out-of-plane flexural displacements, reducing the dispersive $C_{5}T^{5}$ contribution associated
with ZA phonons and enhancing the relative importance of the Debye $C_{3}T^{3}$ term from
linearly dispersing LA and TA modes. At the highest concentrations, the ratio $100|C_{5}|/C_{3}$
approaches very small values, indicating that the intrinsic ZA contribution is almost completely
screened by the adsorbate layer and that the low-energy vibrational spectrum becomes increasingly
dominated by adsorbate excitations. This behavior is consistent with the reduced negative
Gr\"uneisen coefficients \cite{schelling2003thermal} and suppressed transverse acoustic contributions reported for Xe- and
N$_2$-saturated SWCNTs bundles \cite{dolbin2009radial,dolbin2010low}.

The evolution of the effective coefficients $C_{3}$ and $C_{5}$ in the SWCNT--gas 
composites reflects a systematic modification of the low-energy vibrational 
spectrum that is difficult to explain solely by the addition of adsorbate phonon 
modes. As shown in Fig.~\ref{fig:fig_8}, the ratio $|C_{5}|/C_{3}$ decreases non-monotonically. The behavior of the ratio $|C_{5}|/C_{3}$ across the full Ne loading range suggests a
systematic modification of the low-energy vibrational spectrum of SWCNT bundles
that may be difficult to fully reconcile with a purely additive model in which the nanotube host
remains unchanged and the adsorbate contributes an independent vibrational term.
At low coverages (0--50\,scc), the non-monotonic evolution of $|C_{5}|/C_{3}$ reflects
the sequential filling of the energetically preferred interstitial-channel (IC) and
groove (G) adsorption sites\cite{stan1998interstitial,stan2000uptake,ulbricht2002desorption, kostov2003phonons, siber2002phonons}. In this regime,
the effective coefficients follow the scaling relation:

\begin{equation}
|C_{5}| = 1.3\, C_{3}^{0.2}\,,\mathrm{[K^{-2}]},
\label{eq:C5C3_corelationNe}
\end{equation}

indicating an initial renormalization of the ZA flexural modes of the nanotube
bundle. 

Such behavior is difficult to reconcile with a purely additive superposition of nanotube and adsorbate heat capacities, suggesting that adsorbate loading modifies the collective low-energy dynamics of the coupled system. In the intermediate loading regime (50--125 scc), the ratio $|C_5|/C_3$ decreases nearly linearly, indicating that the evolution of the effective coefficients is not governed solely by the increasing adsorbate content. Within the phenomenological framework, this trend points to a progressive restructuring of the low-energy vibrational spectrum of the SWCNT--adsorbate system. One possible interpretation is that increasing adsorbate loading alters the character of low-frequency transverse excitations through changes in intertube and adsorbate--nanotube interactions. This effect may additionally involve a renormalization and partial hybridization of ZA-like nanotube modes with collective vibrations of adsorbed Ne. However, the effective coefficients extracted from heat-capacity measurements do not permit an unambiguous identification of the microscopic mechanism.

The corresponding scaling relation,

\begin{equation}
|C_{5}| = 1.7\, C_{3}^{0.2}\,,\mathrm[{K^{-2}]},
\label{eq:C5C3_corelationNeint}
\end{equation}

captures the modified interplay between $C_{3}$ and flexural excitations $|C_{5}|$ in this
coverage interval. At high Ne coverages (125--150\,scc), the ratio $|C_{5}|/C_{3}$ approaches a plateau,
consistent with the formation of a continuous adsorbate film on the outer surface
(OS) sites of the bundle. Once these sites are saturated, additional Ne atoms behave
almost additively: $C_{5}$ changes only slightly, whereas $C_{3}$ continues to grow.
This saturation suggests that the SWCNT--Ne system has reached a new effective
vibrational state in which the renormalization of the host flexural spectrum is
essentially complete.

A similar trend is observed for SWCNT--Xe and SWCNT--N$_2$ composites, which
exhibit slightly reduced values of $|C_{5}|/C_{3}$ for case of (Xe or N$_2$) particles adsorbed in the grooves compared to pristine SWCNT bundles.
Although the magnitude of the effect differs among adsorbates, the systematic
decrease of $|C_{5}|/C_{3}$ in all cases suggests that adsorption modifies the
flexural spectrum of the host rather than simply adding independent low-energy
adsorbate modes.

Taken together, the non-linear onset at low Ne coverages, the shift of the ratio
near 50\,scc, and the nearly linear decrease in the intermediate regime provide a
strong indication of an interaction-driven contribution $C_{\mathrm{int}}(T)$ that may
renormalize the flexural modes of the SWCNT host. These results suggest that gas
adsorption alters the low-temperature heat capacity of nanotube bundles in a way
that may be difficult to fully capture within a strictly additive model in which the nanotube lattice
remains unchanged and the adsorbate contributes an independent heat capacity.

Such behavior is difficult to fully reconcile with a purely additive scenario in which the adsorbate
contributes only its own low-energy excitations. Localized adsorbate modes or
adsorbate–SWCNT interactions may increase the absolute magnitude of $|C_{5}|$, but
they may not be expected to modify the intrinsic correlation between $C_{3}$ and $C_{5}$ that
characterizes the flexural spectrum of pristine SWCNT bundles. Therefore, the
observed reduction of the ratio $|C_{5}|/C_{3}$ may be indicative of a
renormalization of the host ZA flexural modes.

These results suggest that gas adsorption modifies the low-temperature heat capacity of
SWCNT bundles in a way that may be difficult to fully capture within a strictly additive model in which the nanotube
lattice remains unchanged and the adsorbate contributes an independent heat capacity. The
systematic evolution of $C_{3}$, $C_{5}$, and the ratio $100|C_{5}|/C_{3}$ shows that adsorption
reshapes the low-energy excitation spectrum of the composite SWCNT--gas system. For pristine
bundles, the negative $C_{5}$ coefficient reflects the dispersive contribution of out-of-plane
flexural (ZA) phonons, whose soft quadratic dispersion makes them highly sensitive to mass
loading, boundary conditions, curvature, and intertube coupling.

The concentration dependence of $C_{5}$ is particularly informative for Ne, where the magnitude
of the negative coefficient increases systematically with loading. At the same time, the ratio
$100|C_{5}|/C_{3}$ also decreases non-monotonically with increasing Ne concentration, suggesting
that the dispersive contribution becomes progressively weaker relative to the Debye term as adsorption proceeds. This behavior is consistent with adsorption modifying the dispersive part of the spectrum through changes in local mass distribution, altered van der Waals coupling, and occupation of groove or interstitial sites that influence the mechanical environment of neighboring
nanotubes. Low-energy adsorbate vibrations may also hybridize with the flexural modes of the nanotube framework, producing an effective $C_{5}$ that reflects the combined dispersive response
of the coupled SWCNT--gas system.

Overall, the combined behavior of $C_{3}$, $C_{5}$, and $100|C_{5}|/C_{3}$ provides a strong indication
that adsorption reshapes the low-energy vibrational spectrum of SWCNT bundles through
both direct adsorbate excitations and adsorption-induced modifications of the nanotube ZA flexural modes (Table~\ref{tab:fit_Cp}).

\begin{figure}[htbp]
  \centering
  \includegraphics[width=0.85\textwidth]{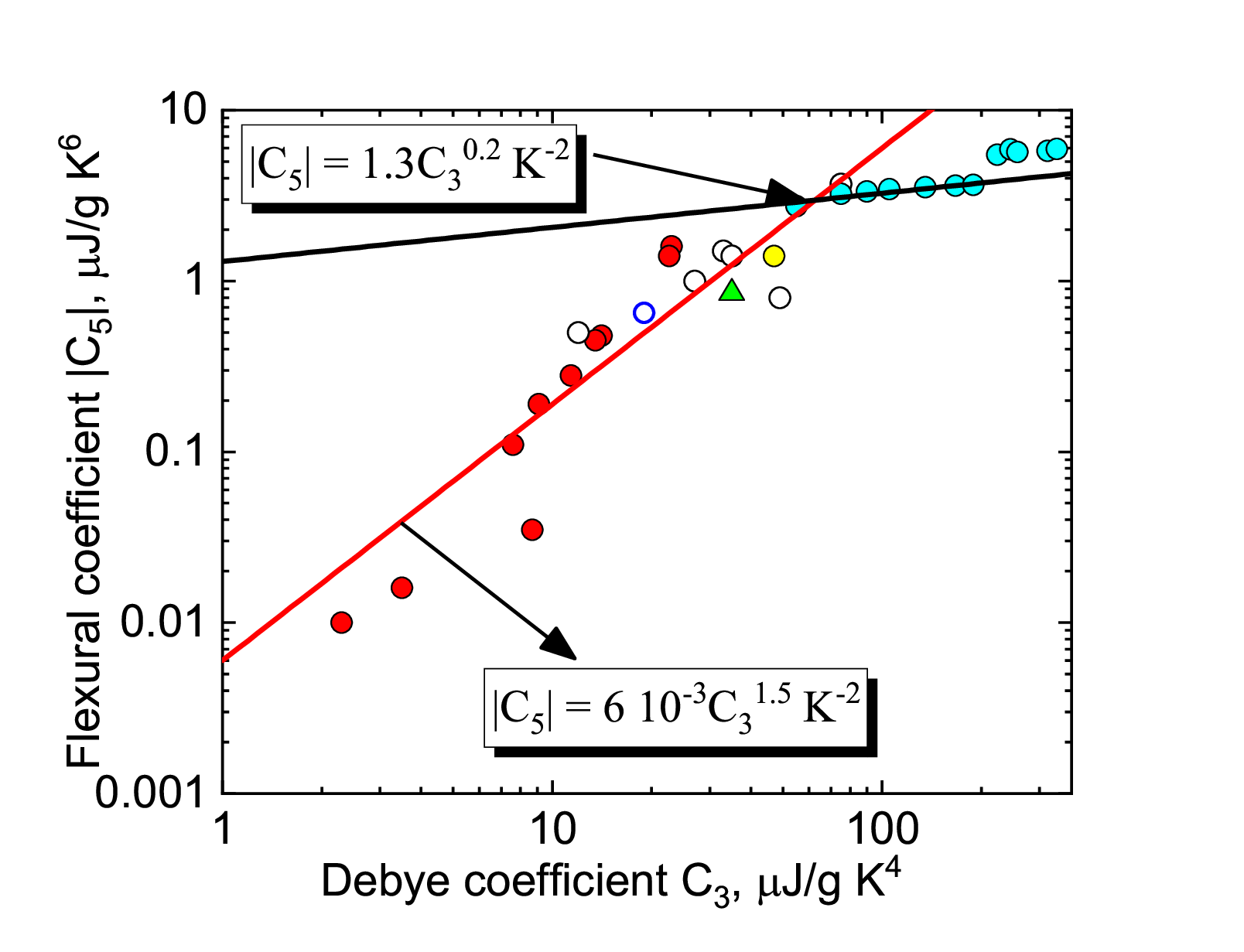}
  
  \caption{Power-law correlation between the flexural ZA coefficient $\lvert C_{5}\rvert$  and the Debye coefficient $C_{3}$ for SWCNTs (\textcolor{blue}{\(\circ\)}) and MWCNTs (\textcolor{red}{\(\bullet\)}), Ne-loaded SWCNT bundles (\textcolor{cyan}{\(\bullet\)}), Xe-loaded SWCNT bundles (\textcolor{yellow}{\(\bullet\)}), and N$_2$-loaded SWCNT bundles \textcolor{green}{\(\blacktriangle\)}). The solid red line represents the intrinsic carbon scaling $\lvert C_{5}\rvert = 6\times10^{-3} C_{3}^{1.5}\,\mathrm{K^{-2}}$. Data for SWCNT--gas composites, shown as Ne-loaded SWCNT bundles (\textcolor{cyan}{\(\bullet\)})  systematically deviate from this trend. These deviations reflect the suppression of the flexural ZA phonon contribution of the SWCNTs-Ne system by adsorbed Ne at high loading of adsorbed atoms.  The solid black line corresponds to the scaling $\lvert C_{5}\rvert = 1.3\times C_{3}^{0.2}\,\mathrm{K^{-2}}$ characteristic of SWCNTs with Ne loading. The fitted $C_{3}$ and $C_{5}$ parameters for SWCNTs-gas systems listed in Table~\ref{tab:fit_Cp}.  }
  \label{fig:fig_9}
\end{figure}

In Figure~\ref{fig:fig_9} is shown the correlation between the flexural ZA coefficient $\lvert C_{5}\rvert$ 
and the Debye coefficient $C_{3}$ for carbon-based nanostructures. 
Pristine SWCNTs and MWCNTs follow a well-defined power-law scaling empirical fit 
\cite{barabashko2025experimental}:
\begin{equation}
\lvert C_{5}\rvert = 6\times10^{-3}\, C_{3}^{1.5}\,\mathrm{K^{-2}}.
\label{eq:C5C3_carbon}
\end{equation}
This relation reflects the intrinsic coupling between linearly dispersing 
acoustic modes (LA, TA) and the quadratic flexural (ZA) branch in 
low-dimensional carbon systems. This “carbon trend’’ represents the universal 
vibrational response of nanotubes in the absence of external perturbations.

The data for SWCNT--Ne composites (blue circles) deviate systematically from 
this trend. For a given value of $C_{3}$, the corresponding $\lvert C_{5}\rvert$ 
values are significantly smaller than expected for pristine nanotubes without renormalization of the ZA flexural-mode dispersion, and the
Ne-loaded points align more closely with a weaker empirical scaling relation given by Eq.~\ref{eq:C5C3_corelationNe}. This behavior indicates a substantial reduction in the relative weight of the 
flexural contribution.

The deviation provides phenomenological signature consistent with modification of
the low-energy dispersive term, potentially including altered ZA
flexural modes of SWCNTs due to Ne adsorption. As Ne atoms populate groove and outer-surface 
sites, they increase the effective mass density of the nanotube walls and 
introduce local pinning potentials. With increasing coverage, the adsorbate 
layer becomes more continuous, stiffening the nanotube surface and altering the 
boundary conditions for out-of-plane vibrations. These effects reduce the 
amplitude and soften the contribution of ZA modes, thereby lowering 
$\lvert C_{5}\rvert$ relative to $C_{3}$.

At the same time, Ne adsorption enhances the Debye-like acoustic contribution: 
the increased mass loading and modified elastic environment lead to a systematic 
growth of $C_{3}$. The combined effect---increase of $C_{3}$ and suppression of 
$\lvert C_{5}\rvert$---shifts the SWCNT--Ne data away from the universal carbon 
scaling and toward the weaker-exponent trend. This behavior is consistent with 
the broader physical picture in which adsorbates progressively screen the soft 
flexural phonons, driving the vibrational spectrum of the composite toward a 
more three-dimensional character described by Eq.~\eqref{eq:C5C3_corelationNeint}.

Thus, the position of the blue circles (SWCNTs with adsorbed Ne) in the figure captures the essential 
physics of Ne adsorption: the flexural ZA contribution becomes increasingly 
suppressed as the adsorbate concentration grows, while the Debye acoustic term 
becomes comparatively more dominant, leading to a characteristic departure from 
the intrinsic scaling law of Eq.~\eqref{eq:C5C3_carbon}.

This interpretation is reinforced by Fig.~9. Pristine SWCNTs and MWCNTs follow 
the intrinsic carbon scaling relation $|C_{5}| \propto C_{3}^{1.5}$, whereas 
Ne-loaded composites at high coverage transition to a distinct scaling 
$|C_{5}| \propto C_{3}^{0.2}$. Such a change in the scaling exponent is difficult to fully reconcile with 
adding an independent adsorbate contribution alone; instead, it may reflect a 
modified interplay between the linear and flexural-like modes of the host. 
Therefore, the effective parameters $C_{3}^{\mathrm{eff}}$ and $C_{5}^{\mathrm{eff}}$ 
already encode the renormalization of the SWCNT bundle’s flexural dispersion induced 
by the adsorbed atoms. We interpret these changes as a thermodynamic signature of 
an effective interaction term $C_{\mathrm{int}}(T)$ in Eq.~\eqref{eq:total_heat_capacity}, without attempting to extract 
it explicitly or relying on a specific microscopic model of the adsorbate.

The behavior of the SWCNT-Xe series in Fig.~\ref{fig:fig_9} is fully consistent with measurements of 
one-dimensional Xe chains adsorbed in SWCNT grooves. Barabashko \textit{et al.} 
\cite{barabashko2023experimental} demonstrated that the heat capacity $C_{\mathrm{ads}}$ of Xe chains 
is described by a vibrational contribution $C_{\mathrm{ads}}^{\mathrm{model}}$ \cite{siber2002phonons, manzhelii2019phonon} below $8\,\mathrm{K}$. 
This interpretation is also in quantitative agreement with other established theoretical 
models of rare-gas adsorption on carbon substrates. Calculations by Siber \cite{siber2002phonons}, 
Manzhelii el al. \cite{manzhelii2019phonon,manzhelii2025localized}, and Syvokon et al. \cite{syvokon2023collective} for Xe atoms in groove and surface sites of SWCNT bundles 
show that the adsorbate vibrational spectrum $C_{\mathrm{ads}}^{\mathrm{model}}(T)$ closely 
matches the experimentally reconstructed $C_{\mathrm{ads}}^{\mathrm{exp}}(T)$. In particular, 
the linear-in-$T$ contribution associated with longitudinal acoustic modes in the 
Xe chains exhibits excellent agreement between theory and experiment \cite{bagatskii2013experimental,siber2002phonons}. These models 
consistently indicate negligible renormalization of the nanotube ZA dispersion, 
implying that at the low coverages studied here Xe acts primarily as an inert mass 
load rather than a source of additional low-energy collective modes. As a result, 
Xe adsorption produces only mass loading and a very weak modification of intertube 
coupling, while leaving the dispersive ZA modes of the nanotube bundle only slightly 
perturbed. This agrees with our observation that the $|C_{5}|/C_{3}$ 
for SWCNT--Xe decrease only marginally and remains close to the pristine SWCNT 
trend. Consequently, the interaction term $\Delta C_{\mathrm{int}}(T)$ for Xe is negligible 
within experimental accuracy, consistent with Xe inducing only a minimal reshaping 
of the nanotube vibrational spectrum.

Our experimental results follow this theoretical expectation: the effective 
coefficient $C_{5}$ for SWCNT--Xe increases only slightly and remains close to the 
pristine SWCNT correlation given by Eq.~\ref{eq:C5C3_carbon}, while the reconstructed Xe curves preserve the original 
$C_{3}$--$C_{5}$ scaling of the host. Consequently, Xe adsorption in the groove sites 
causes only minimal reshaping of the nanotube vibrational spectrum, and the 
interaction term $\Delta C_{\mathrm{int}}(T)$ for Xe effectively vanishes within 
experimental accuracy.

\subsection{The coefficient $C_{5}$ as a phenomenological fingerprint of adsorption for SWCNTs-gas system}\label{subsec5}

Flexural modes of SWCNTs are particularly sensitive to the gas-adsorbed 
environment because their frequencies depend not only on the bending rigidity 
of the nanotube walls but also on mass density, boundary conditions, curvature, 
and intertube coupling. Consequently, adsorption of even weakly bound atoms or 
molecules can produce measurable changes in the low-temperature thermodynamic 
response.

It is therefore useful to distinguish two effects of adsorption. The first is 
the direct contribution of the adsorbate degrees of freedom to the heat 
capacity. The second is the modification of the vibrational spectrum of the 
nanotube bundle caused by the additional mass and by adsorbate--nanotube and 
adsorbate--adsorbate interactions. The experimental results indicate that both 
effects are important, with their relative contributions depending strongly on 
the adsorbate species and concentration.

The evolution of the effective coefficients $C_{3}$ and $C_{5}$ in the SWCNT--gas
composites suggests that adsorption modifies the low-energy vibrational spectrum
of the nanotube bundles in a manner that may not be attributable solely to the addition
of adsorbate-specific modes. As shown in Fig.~8, the absolute magnitude $|C_{5}|$
increases with coverage, whereas the relative dispersive weight $|C_{5}|/C_{3}$
decreases. This behavior suggests that the Debye-like linear contribution $C_{3}$
grows more rapidly than the flexural-like contribution $C_{5}$, which is difficult to fully reconcile
with a purely additive scenario. Additional localized adsorbate modes
would increase $C_{3}$ and $|C_{5}|$ without altering the intrinsic correlation between $C_{3}$
and $C_{5}$. The observed reduction of $|C_{5}|/C_{3}$ therefore indicates a
renormalization of the ZA flexural dispersion for the SWCNTs.

This conclusion is supported by the scaling analysis. Pristine SWCNTs and
MWCNTs follow the intrinsic carbon relation

\begin{equation}
|C_{5}| \propto C_{3}^{1.5},
\label{eq:C5C3_carbonnanotubes}
\end{equation}

reflecting the characteristic interplay between linear acoustic and flexural ZA-like
modes. In contrast, Ne-loaded composites at high coverage exhibit a distinct scaling

\begin{equation}
|C_{5}| \propto C_{3}^{0.2},
\label{eq:C5C3_carbon_and_Ne}
\end{equation}

Such a substantial change in the scaling exponent is difficult to fully reconcile with simply adding an
independent adsorbate heat-capacity term; it likely reflects a modification of the effective
dispersion in the nanotubes.

Crucially, the data for Xe-loaded bundles, N$_2$-loaded bundles, and for Ne at low
concentrations---corresponding to the filling of groove sites---continue to lie on the
intrinsic carbon scaling line with $\alpha \simeq 1.5$. This result suggests that, below a certain impurity concentration, the effective coefficients $C_3^{\rm eff}$ and $C_5^{\rm eff}$ evolve in a manner broadly consistent with the response of the host nanotube bundle. Within the sensitivity of the present analysis, no distinct signature of an interaction-induced contribution to the heat capacity can be identified. Consequently, any coupling-related contribution $C_{\rm int}(T)$ is likely to remain below the resolution of the available calorimetric data in this concentration range. Only above this crossover loading does the system deviate from
intrinsic scaling, indicating that the host and adsorbate contributions can no longer
be treated as independent. In this high-coverage regime, the emergence of the new
exponent $\alpha \simeq 0.2$ signifies a renormalization of the effective ZA-mode
dispersion of the nanotube bundle.

Thus, the effective parameters $C_{3}^{\mathrm{eff}}$ and $C_{5}^{\mathrm{eff}}$ already
encode the interaction-driven modification of the flexural spectrum induced by
adsorption. We interpret these changes for Ne as a thermodynamic signature of a non-zero
interaction term $C_{\mathrm{int}}(T)$, without attempting to extract it explicitly or
relying on a specific microscopic model of the adsorbate.

\subsubsection{ Mass loading and modification of flexural modes}

Although carbon nanotubes are not chain structures in the sense of Lifshitz’s original 
model, they exhibit low–energy quasi–flexural phonon modes with the similar quadratic 
dispersion $\omega \propto q^{2}$ curvature predicted for layered and chain systems~\cite{gospodarev2017phonon}. 
Studies of flexural phonon dynamics in nanotubes, together with analyses of the ZA 
branch in graphene, are consistent with this behavior\cite{gospodarev2017phonon,kosevich2006crystal,taheri2021importance,barabashko2025experimental}. 
Wave propagation in carbon nanotubes (CNTs), described using nonlocal Euler–Bernoulli 
and Timoshenko beam models by Wang~\cite{wang2005wave}, shows that the group velocity :
\begin{equation}
v_{\mathrm{ZA}}\propto 
\left( \frac{\kappa_{\mathrm{eff}}}{\mu_{\mathrm{eff}}} \right)^{1/2},
\end{equation}
and near the Brillouin–zone center the ZA phonon branch is governed by
\begin{equation}
\omega_{\mathrm{ZA}}(q) \propto 
\left( \frac{\kappa_{\mathrm{eff}}}{\mu_{\mathrm{eff}}} \right)^{1/2} q^{2},
\end{equation}
where $\kappa_{\mathrm{eff}}$ and $\mu_{\mathrm{eff}}$ are the effective bending rigidity and mass per unit length, respectively.

In the presence of intertube coupling or residual tension, the flexural branch acquires 
a linear long–wavelength component, leading to the upward curvature described by the 
relation:
\begin{equation}
\omega^{2} = v^{2} q^{2} + D q^{4}, \qquad D = \frac{\kappa_{\mathrm{eff}}}{\mu_{\mathrm{eff}}},
\label{eq:dispersion_upward}
\end{equation}
which is formally analogous to the dispersion law derived by Kosevich for a stretched 
linear chain~\cite{kosevich2006crystal}. In this framework, the low–energy phonons exhibit a 
crossover from the linear acoustic regime to the quadratic flexural dispersion regime and the thermodynamic signature of this transition is the 
appearance of a negative $C_{5}$ term in the heat capacity.

A useful parameter that characterizes the interplay between the linear and quadratic 
parts of the flexural dispersion is the crossover temperature $T_{\times}$ defined as the temperature at which the magnitudes of the two fitted contributions 
become equal,
\begin{equation}
C_{3}T_{\times}^{3}=|C_{5}|T_{\times}^{5}.
\label{eq:balance}
\end{equation}
Equation~\eqref{eq:Tcross} provides a characteristic $T_{\times}$ marking
the onset at which the dispersive correction associated with the flexural branch becomes 
comparable to the Debye-like acoustic $T^{3}$ term. A larger $T_{\times}$ therefore indicates 
that the system remains dominated by the linear acoustic regime over a broader 
temperature interval, and that the quadratic flexural contribution becomes 
thermodynamically relevant only at higher temperatures.

In the context of the dispersion relation introduced in Eq.~\eqref{eq:dispersion_upward}, the ratio 
$|C_{5}|/C_{3}$ serves as a sensitive thermodynamic indicator of the balance between 
the linear and quadratic components of the ZA phonon branch. Adsorbed particles such 
as Xe, Ne, and N$_2$ modify this balance primarily through mass loading, which 
increases the effective mass per unit length $\mu_{\mathrm{eff}}$, and through weak 
pinning at groove sites and stronger pinning at outer–surface sites, which alters the 
effective bending rigidity $\kappa_{\mathrm{eff}}$. Within the Lifshitz–Kosevich 
framework \cite{kosevich2006crystal}, these changes reduce the parameter 
$D$, thereby shifting the crossover frequency $\omega_{\times}$ and increasing the 
corresponding crossover temperature
\begin{equation}
T_{\times}=\left(\frac{C_{3}}{|C_{5}|}\right)^{1/2},
\label{eq:Tcross}
\end{equation}
at which the magnitudes of the two fitted contributions become equal 
($C_{3}T_{\times}^{3}=|C_{5}|T_{\times}^{5}$).

This provides a natural explanation for the experimentally observed reduction of the 
ratio $|C_{5}|/C_{3}|$ in Xe–loaded nanotube bundles. The acoustic contribution 
($C_{3}$ term) increases due to mass loading and softening of long–wavelength modes, 
while the flexural contribution ($C_{5}$ term) remains negative but becomes relatively 
less significant within the same temperature interval. In other words, adsorption does 
not simply enhance the flexural response; instead, it shifts the linear–quadratic 
crossover of the ZA branch to higher temperatures, reducing the relative weight of the 
quadratic component. This behavior preserves the overall "carbon–based trend" described by eq~\ref{eq:C5C3_carbonnanotubes} observed 
for nanotube bundles and reflects the general consequence of adsorption: an increase 
in $\mu_{\mathrm{eff}}$ and a modification of the effective restoring forces and boundary 
conditions that govern the flexural spectrum. This behavior indicates that Xe adsorption does not 
simply enhance the flexural modes but changes the relative weight of the linear and 
quadratic components of the ZA dispersion, preserving the overall carbon–based trend 
observed for nanotube bundles.

The most general consequence of adsorption is an increase in the effective mass 
associated with the Ne adsorbed on the nanotube surface. Adsorption therefore modifies 
the flexural spectrum primarily through the increase of $\mu_{\mathrm{eff}}$. 
At the same time, interactions between adsorbates and the nanotube surface, as 
well as between neighboring nanotubes, may alter the effective restoring forces 
and boundary conditions.
SWCNTs-Ne systems provide particularly useful evidence in this respect. Ne has no internal rotational degrees of freedom, yet Ne adsorption produces systematic changes in 
both $C_{3}$ and $C_{5}$. Thus, the observed changes may not be attributable solely 
to molecular rotation or libration; they suggest that adsorption modifies 
the low-energy thermodynamics of the nanotube--adsorbate system interaction itself. At the same time, the experimental results do not permit the changes in $C_{5}$ 
to be interpreted quantitatively as a direct change in the bending rigidity or 
ZA-phonon dispersion. The fitted coefficient is obtained from the total heat 
capacity and therefore characterizes the effective low-energy response of the 
SWCNTs-gas system.

\subsubsection{ Discussion of Overall physical picture of modification of flexural modes}

The comparison between different SWCNTs with adsorbed species (Xe, Ne, N$_2$, CH$_4$) 
suggests that adsorption modifies the heat capacity through two complementary 
mechanisms: mechanical perturbation of the nanotube vibrational spectrum and the 
introduction of new adsorbate degrees of freedom. Xe provides perhaps the clearest case of 
the former because it has no internal rotational modes, while Ne illustrates the effect of 
progressive population of adsorption sites, while N$_2$ and CH$_4$ highlight the 
additional role of molecular dynamics. For this reason, the coefficients $C_{3}$ and 
$C_{5}$ obtained for molecular adsorbates should be interpreted as effective parameters 
of the SWCNT--gas system rather than as intrinsic properties of pristine SWCNTs.

The systematic evolution of $C_{5}$ with adsorbate type and concentration indicates 
that this coefficient serves as a useful phenomenological indicator of 
adsorption-induced changes in the low-energy excitation spectrum. For pristine SWCNT 
bundles, the negative $C_{5}$ term is naturally associated with flexural ZA-mode 
dispersion \cite{barabashko2025experimental}. Upon adsorption, however, its magnitude reflects the combined influence 
of modified flexural modes and additional dispersive excitations introduced by the 
adsorbate. The coefficient should therefore not be regarded as a direct measure of 
bending rigidity, ZA-phonon scattering, or the strength of adsorbate--nanotube 
coupling. Instead, its value lies in providing a reproducible experimental parameter 
with which different SWCNT--gas systems can be compared. The systematic evolution of 
$C_{5}$ with Ne coverage and its distinct behavior for Xe, N$_2$, and CH$_4$ support the view 
that the observed changes are governed primarily by the adsorbate environment rather than by 
random deviations from the polynomial description. A quantitative assignment of 
$C_{5}$ to individual vibrational branches would require explicit calculations of the 
full coupled SWCNT--adsorbate spectrum.

The results can be summarized within a coupled-system picture in which the low-energy 
excitations arise from three interrelated components: (i) the vibrational spectrum of 
the SWCNT bundle modified by adsorption, (ii) intrinsic translational and, for 
molecular adsorbates, rotational or librational excitations of the adsorbate, and 
(iii) coupling between the two subsystems. In this framework, adsorption does not 
simply add an independent heat-capacity contribution; it alters the local mass 
distribution, mechanical boundary conditions, intertube van der Waals interactions, 
and, at sufficiently high coverage, the structural order of the adsorbate layer.

This interpretation is supported by the low-temperature heat capacity of rare-gas films studied by Menges and v.~Löhneysen~\cite{menges1991specific}. 
Their measurements revealed substantial low-energy contributions that are not readily 
described by an ideal crystalline phonon spectrum; in particular, a nearly 
linear-in-$T$ term was observed \cite{menges1991specific}. 
Thus, adsorbed rare-gas species possess a broad spectrum of low-energy excitations 
arising from heterogeneous local environments. In SWCNT bundles, such excitations 
may coexist and interact with the low-energy flexural modes of the nanotube lattice.

The experimental results therefore support the following physical picture. Gas 
adsorption does not simply add an independent heat-capacity term to an otherwise 
unchanged SWCNT lattice. Instead, the composite vibrational density of states can be 
represented schematically as

\begin{equation}
g_{\mathrm{tot}}(\omega)
= g_{\mathrm{SWCNT}}^{*}(\omega)
+ g_{\mathrm{gas}}(\omega)
+ g_{\mathrm{int}}(\omega),
\label{eq:g_tot}
\end{equation}

where $g_{\mathrm{SWCNT}}^{*}$ denotes the nanotube spectrum renormalized by 
adsorption, $g_{\mathrm{gas}}$ represents intrinsic adsorbate excitations, and 
$g_{\mathrm{int}}$ accounts phenomenologically for their coupling. The negative 
$C_{5}$ term thus provides a sensitive signature of the dispersive low-energy sector, 
and its systematic modification upon adsorption suggests that the low-energy 
vibrational response of SWCNT bundles is strongly influenced by the adsorbate 
environment. However, the magnitude of $C_{5}$ alone does not distinguish between 
changes in intrinsic flexural modes and the emergence of adsorbate excitations; this 
typically requires joint consideration of $C_{5}$, adsorbate concentration, the overall heat 
capacity, and the ratio $C_{\mathrm{SWCNT}}/C_{\mathrm{SWCNT+gas}}$.

The combined behavior of Xe, Ne, N$_2$, and CH$_4$ provides evidence that gas adsorption can renormalize the low-energy vibrational response of 
SWCNT bundles. Atomic adsorbates primarily induce mechanical perturbations, whereas 
molecular species introduce additional dynamical degrees of freedom. Low-temperature 
calorimetry therefore offers a sensitive macroscopic probe of adsorption-induced 
modifications of the low-energy spectrum. A more definitive microscopic description will 
require calculations that explicitly include adsorption-site distributions, intertube 
coupling, adsorbate coverage, and the full coupled phonon--adsorbate spectrum, which 
could establish quantitatively how adsorption renormalizes the ZA dispersion and 
determine the relative contributions of nanotube and adsorbate excitations to the 
experimentally observed $C_{3}$ and $C_{5}$ coefficients.

\section{Conclusions}\label{sec4}

The present study suggests that adsorption of Xe, Ne, N$_2$, and CH$_4$ systematically modifies the low-temperature thermodynamic response of SWCNT bundles and is consistent with changes in the effective low-energy vibrational spectrum. Across all systems, the heat capacity remains well described by the 
phenomenological form

\[
C(T)=C_{3}T^{3}+C_{5}T^{5},
\]

which enables quantitative comparison of the "acoustic" ($C_{3}$) and "dispersive" ($C_{5}$) terms. 
Gas adsorption modifies both coefficients in a manner that reflects the coupled 
response of the nanotube lattice and the adsorbate layer.

At low coverage, weakly bound atoms or molecules perturb the nanotube spectrum 
only slightly, and signatures of intrinsic flexural (ZA) modes remain visible. With 
increasing coverage, mass loading and changes in local boundary 
conditions enhance the acoustic contribution $C_{3}$, while 
reducing $|C_{5}|/C_{3}$. This reduction is strongest for heavy adsorbates and for dense 
films. This trend is consistent with progressive suppression of the dispersive contribution characteristic of pristine bundles, potentially associated with surface pinning and enhanced intertube coupling.

Molecular adsorbates introduce additional complexity: rotational, librational, and orientational excitations contribute to the low-energy heat capacity and are reflected in the effective coefficient $C_{5}$ obtained from the phenomenological fit. Thus, the evolution 
of $C_{5}$ reflects not only modification of the nanotube flexural modes but also the 
emergence of low-energy adsorbate dynamics.

Taken together, the results support a unified physical picture in which adsorption reshapes 
the balance between acoustic and flexural excitations in SWCNT bundles through mass 
loading, altered boundary conditions, intertube coupling, and molecular dynamics. 
The sensitivity of the coefficients $C_{3}$ and $C_{5}$ to adsorbate type and 
concentration provides a useful phenomenological fingerprint of these changes. 
Low-temperature calorimetry therefore offers a sensitive macroscopic probe of adsorption-induced modifications of the effective low-energy vibrational spectrum and provides a controlled 
route for tuning and investigating the low-energy vibrational properties of 
low-dimensional carbon systems. 
The results further indicate that the negative coefficient $C_{5}$ should be regarded as a phenomenological fingerprint of the collective low-energy excitations of the coupled SWCNT-adsorbate system rather than as a unique measure of intrinsic ZA phonons of pristine nanotubes. The possible connection to transverse vibrational modes of nanotube bundles remains a physically motivated interpretation that requires additional experimental and theoretical verification.

\backmatter

\bmhead{Acknowledgements}
This work was partly supported by the National Research Foundation of Ukraine (Grant 2023.03/0012) and the National Academy of Sciences of  Ukraine (№ 0122U001504).


\bibliography{sn-bibliography}

\end{document}